\documentclass[11pt]{article}
\usepackage{amsmath,amsfonts, amssymb,graphicx,geometry,authblk,color,soul,comment,hyperref,amsthm,wrapfig} 
\usepackage{subcaption}
\usepackage[italicdiff]{physics}
\title{Learning constitutive relations from experiments: \\
1. PDE constrained optimization}
\author[1]{Andrew Akerson}
\author[2]{Aakila Rajan}
\author[2]{Kaushik Bhattacharya}
\affil[1]{Reality Labs Research, Meta Platforms, Inc., Redmond, WA 98052, USA}
\affil[2]{Division of Engineering and Applied Science, California Institute of Technology, Pasadena, CA 91125, USA}
\date{\today}

\begin{document}
\maketitle

\begin{abstract}

We propose a method to accurately and efficiently identify the constitutive behavior of complex materials through full-field observations.  We formulate the problem of inferring constitutive relations from experiments as an indirect inverse problem that is constrained by the balance laws.   Specifically, we seek to find a constitutive behavior that minimizes     the difference between the experimental observation and the corresponding quantities computed with the model, while enforcing the balance laws.   We formulate the forward problem as a boundary value problem corresponding to the experiment, and compute the sensitivity of the objective with respect to model using the adjoint method.  The resulting method is robust and can be applied to constitutive models with arbitrary complexity.  We focus on elasto-viscoplasticity, but the approach can be extended to other settings.  In this part one, we formulate the method and demonstrate it using synthetic data on two problems, one quasistatic and the other dynamic.
%
\end{abstract}

\section{Introduction}

Continuum mechanics is an approach to solving problems involving complex phenomena directly at the scale of applications  (e.g., \cite{gurtin_1970,billington_1981}).  It exploits universal laws of physics: the balance of mass, momenta, energy as well as the second law of thermodynamics.  However these equations are not closed and a constitutive relation that characterizes the material is necessary to do so.  This constitutive relation is typically obtained empirically by conducting experiments.   However, we cannot measure the constitutive relation directly and it has to be obtained through the solution of an inverse problem.   In fact, we cannot even directly measure the quantities like stress, strain, strain rate and energy density that comprise the constitutive relation.  Instead, they have to be inferred from quantities like displacements and total forces that can be measured in the laboratory.  Thus, the problem of inferring constitutive relations from experiments, and thereby completing the continuum mechanical formulation, requires the solution of an indirect inverse problem.

The classical approach is to design an experimental setup that is consistent with either uniform states of stress and strain/strain-rate (e.g., uniaxial tension in solids or shear rheometers in complex fluids) or a universal solution (e.g., torsion in solids or viscometric flow in complex fluids); these enable a semi-analytic solution to the indirect inverse problem (e.g. \cite{billington_1981}).  There are a number of difficulties with this approach.  First, we obtain information about the constitutive relation only in simple, idealized states of stress (history) while we seek to deploy them in more complex situations.  For example, we may obtain only uniaxial tensile data while we may want to use the model in multiaxial, non-proportional loading scenarios.  Second, one has to conduct a large number of tests since each test can only provide a limited amount of information on the constitutive behavior.   For example, we may have to repeat a test at various strain rates, or with specimens with different orientation.  Third, one may need sterile conditions with precise alignment (e.g., plate impact) to obtain the desired state.  This limits the number of tests that one can conduct.  Together, these and other shortcomings contribute significantly to the uncertainty and limit the fidelity of the constitutive relations that can be obtained from experiments.

It is also classical to postulate {\it a priori} an explicit formula for the constitutive relation with a few constants, and then use above-described experiments to fit the constants  (e.g. \cite{billington_1981}).   Since it is common practice to use a very few constants, the limited experimental information to sufficient to fit them.  However, the ability of explicit formulas with a few constant to represent complex behavior is limited.  Even in the setting of isotropic, incompressible hyperelasticity, we have a still-evolving menagerie of models (e.g. \cite{melly,khaniki}).   It is therefore natural to try to go beyond hypo-parametrized explicit formulas to hyper-parametrized neural networks and other approximations that have proven to be enormously successful in a variety of fields.  However, this requires significant amounts of data, much more than the traditional approach can deliver.

There is a closely related issue.   In history dependent and structured continua, one also has to postulate {\it a priori} the internal or state variables that describe the history dependence/internal structure \cite{billington_1981}.  For example, we introduce plastic strain, accumulated plastic strain and a kinematic hardening variable as internal variables in plasticity.  However, the choice of internal variables is unclear in complex anisotropic situations like in composite media or shape memory alloys \cite{horstemeyer}.  Instead, one would like to infer these directly from experimental measurements, further increasing the need for empirical data.

The recent decade has seen a revolution in experimental methods primarily driven by the growth of full-field observations techniques.  Digital image correlation (DIC)  \cite{chu1985applications, sutton2009image,hild}, where one infers deformation by comparing images of a complex (speckle) pattern imprinted on a surface taken before and after deformation.  The method has been extended to stereo DIC to obtain out-of-plane deformation \cite{sutton_2008}, the high resolution setting using the scanning electron microscope \cite{kammers_digital_2013}, to dynamics  \cite{ravindran2023three} and to digital volume correlation \cite{buljac_digital_2018,yang_augmented_2020}.  DIC enables the measurement of the entire deformation and strain field on a surface.  This opens the intriguing possibility of breaking away from homogeneous (or idealized) states: instead, one can probe many different strain paths in a single test by working with specimens with complex geometries.  However, this requires one to solve the difficult indirect inverse problem.  While one can measure the strain field, it is limited to a single surface.  More importantly, it is not possible to measure the stress field, and instead, we typically can only measure  the total reaction force on a surface.

Another rapidly evolving technique that provides full-field information is high energy diffraction microscopy (HEDM) that uses high energy synchrotron radiation and diffraction to obtain the crystal structure on a three-dimensional voxelated volume \cite{hedm}.  Comparing the obtained crystal structure with the stress-free structure gives us the lattice strain on a voxelated volume.  One can convert this to stress by using the elastic modulus of the material.  However, there are two issues.  First, one only obtains the lattice strain, and not the total strain.  Therefore, one still has to solve an indirect inverse problem in inelastic phenomena.   Second, this is averaged over the volume of the voxel with some unknown kernel, and therefore subject to errors.  For example, the resulting stress fields obtained by this approach are not equilibrated \cite{zhou_2022}.

A variety of approaches have been proposed to solve this indirect inverse problem.  The key idea is to use the balance laws, and specifically equilibrium in some form.   We review a few contributions to provide context to our work, without attempting a comprehensive review of this rapidly growing literature.  One approach is to use a combination of model reduction and Gaussian process regression to obtain the parameters associated with a particular constitutive relation (see \cite{wang_metamodeling_2021, wang2022gaussian} for the use of this approach in viscoelastic materials).   It is possible to combine Bayesian uncertainty quantification with such approaches (see \cite{wu_parameter_2021} on toughness).  While such approaches are effective in identifying a small number of parameters, they scale exponentially with the number of parameters and are prohibitively expensive for complex materials or neural network representations.   While we do not address HEDM in this paper, the ideas presented may be extended to this technique.

Miller, Dawson and collaborators pioneered the concurrent use of finite element analysis with given constitutive relations and experimental observations using high energy x-ray diffraction with the goal of understanding the relationship between single crystal elastic modulus and hardening laws and heterogeneity in stress distribution in polycrystals.\cite{efstathiou_method_2010,obstalecki_quantitative_2014,wielewski_methodology_2017,carson_characterizing_2017}.  However, they do not seek to obtain the full constitutive relationship.  

An emerging approach is to use a physics informed neural network (PINN) where the constitutive relation is represented as a neural network and the (failure of the fields to satisfy) balance laws are used as a part of the objective (loss function) to be minimized as a part of the fitting  \cite{hamel_calibrating_2022, eghtesad_nn-evp_2023, upadhyay_physics-informed_2023}.  The balance laws are not enforced exactly in this approach.

In the the virtual field method (VFM) (for example,  \cite{grediac_virtual_2006,pierron_virtual_2012,kramer_scherzinger_wm_implementation_2014,  marek_sensitivity-based_2017, kim_finite_2021}), the balance laws are integrated against a set of ``virtual fields'' to obtain a system of equations for the unknown constants in a postulated constitutive relation.  The system is linear when the constitutive relation is linear in the constants as in the case of hyperelasticity, but is typically solved by linearization otherwise.  The application of the method requires one to specify kinematically admissible virtual fields, and one may taken them to be the measured strain field.  However, the strain field is typically measured on a part of the boundary of the specimen, and has to be extrapolated to obtain it over the entire domain. Further, one needs as many independent full field measurements as there are unknown constitutive relation even for complex domains.  Finally, only a finite dimensional approximation of the balance laws is enforced.

A widely used approach is the finite element updating (FEMU) (for example, \cite{kavanagh_finite_1971,pagnacco_parameter_2013,mathieu_estimation_2015,neggers_simultaneous_2019,henriques2022inverse, sakaridis2024post, kosin2024parameter, rojicek2021study, conde2023design}  ).
The idea is to formulate the problem as an inverse method which is solved iteratively by using a finite element method to solve the forward problem, and then to update the model parameters using a numerically computed sensitivity.  This has been applied to a variety of history dependent phenomena and can be integrated with the inverse problem of digital image correlation \cite{mathieu_estimation_2015}.   An important consideration here is that the sensitivity, or the gradient of the objective with respect to the objective, is computed numerically.  While a number of ideas can be used to speed this up, this approach scales poorly with the number of material parameters.   A closely related idea is to formulate the problem as a PDE-constrained optimization problem where the balance laws act as the PDE constraint, and then use the adjoint method to compute the sensitivity.  This has been applied to elasticity \cite{oberai_solution_2003}, viscoelasticity \cite{constantinescu_identification_2001} and Norton-Hoff viscoplasticity \cite{bonnet_inverse_2005}.  

In this work, we build on the last two lines of work, and propose a method to accurately and efficiently identify the constitutive behavior through experiments.  We formulate this as an indirect inverse problem that is constrained by the balance laws (PDE constraint).  The objective that we seek to minimize is the difference between the experimental observation and the corresponding quantities computed using the constitutive model.  We then formulate the forward problem as a boundary value problem corresponding to the experiment, and the problem of computing sensitivity of the solution to the parameters as an adjoint problem.  This is a partial differential equation that is linear in space and quasilinear in time.  The adjoint equation has widely been used in optimal design in mechanics (e.g., \cite{Bendsoe2004,Akerson2023}).  In this work, we implement both the forward and the adjoint problem using finite element approximation.   The cost of solving the adjoint problem is that of a single iteration of the forward problem, and the parameters can be updated at little cost from the solution of the adjoint problem.  Thus, the core computation of the method is independent of the number of parameters, and thus suited for complex models with numerous parameters (including neural networks).  

We formulate and demonstrate the method for both quasistatic and dynamic experiments with synthetic data in the current Part 1.  We extend the formulations to include contact in Part 2 and demonstrate it using experimental data from a dynamic indentation test.  We demonstrate the method for models formulated as neural networks in Part 3.    Together, these show that the method is robust and can be adapted for various experimental situations.   While we focus on elasto-viscoplasticity, the ideas are valid for any local constitutive relation.

The paper is organized as follows.  We present the method in Section \ref{sec:form}.  We do so for an arbitrary inelastic material described by a rate-dependent internal variable theory in Section \ref{sec:gen_form}, and for a J2 elasto-viscoplastic model with yield in Section \ref{sec:J2_plastic}.  We overview the numerical method in Section \ref{sec:num} with details provided in the appendix.  We demonstrate and validate the method using synthetic data on two examples in Section \ref{sec:dem}: a thick plate with a hole in quasistatic compression in Section \ref{sec:syn_qs} and an extended dynamic impact test in Section \ref{sec:syn_dyn}.  We conclude with some comments in Section \ref{sec:conc}.

\section{Formulation and method} \label{sec:form}

We first present our method for a general history-dependent constitutive relation, and then specialize to elasto-viscoplasticity.

\subsection{Internal variable theory} \label{sec:gen_form}


\subsubsection{Governing equations} We consider an open, bounded domain $\Omega \subset \mathbb{R}^n$ occupied by a solid body undergoing a deformation with displacement $u(x,t)$ over time interval $(0, T)$. We assume the body is composed of a material of density $\rho$, and one whose constitutive behavior is described by a set of internal parameters $\xi := \{ \xi_i \}, \ i = 1,\dots,N_\xi$ that evolve according to a Perzyna-type kinetic law:
\begin{equation}
\begin{aligned}
    \sigma  = S(F, \xi; P), \\
    \dot{\xi} = R(F, \xi; P) 
\end{aligned} \label{eq:gen_cons_law}
\end{equation}
where $\sigma$ is the first Piola-Kirchoff stress, $F$ is the deformation gradient, and $S$ and $R$ are constitutive functions parametrized by material parameters $P := \{ P_i\}$, $i = 1,\dots, N_p$.    The deformation of the body is then governed by the equations
\begin{equation} \label{eq:gen_eqil}
\begin{aligned}
    &\rho \ddot{u} - \nabla \cdot \sigma = b &&\text{ on } \Omega, \\
    &\sigma  = S(\nabla u, \xi; P), && \text{ on } \Omega \\
    &\dot{\xi} = R(\nabla u, \xi; P) && \text{ on } \Omega, \\
    &\sigma n = f &&\text{ on } \partial_f \Omega,    \\
    &u=u_0 &&\text{ on } \partial_u \Omega,    \\
    &u |_{t = 0} = \dot{u} |_{t = 0} = \xi |_{t = 0} = 0, &&
\end{aligned}
\end{equation}
where $b$  is the applied body force, $f$ is the applied (dead) traction on part $\partial_f \Omega$ of the boundary while $u_0$ is the applied displacement on the complement $\partial_d \Omega$ of the boundary.  Note that $\partial_f \Omega$ typically includes regions that are traction-free.  We assume quiescent initial conditions.

\subsubsection{Indirect inverse problem of parameter identification} 
We are given experimental data, usually displacements at certain instances of time on just a portion of the boundary of the domain (some subset of the traction-free portion of $\partial_f \Omega$) and some overall force or moment acting on some part of the boundary (some subset of $\partial_u \Omega$).  We label the experimental data $D^\text{exp} := \{u^\text{exp}, M^\text{exp}\}$. Here, $u^{exp}$ is the set of partial displacement measurements, and $M^\text{exp}$ is a collection of macroscopic measurable quantities which could consist of net loads or averaged strains. The goal is to find the parameter set $P$ such that the modeled trajectory and computed macroscopic quantities from solving \eqref{eq:gen_eqil} match the experimental data $D^\text{exp}$. We write this as an optimization problem
\begin{equation} \label{eq:gen_opt}
\begin{aligned}
    \inf_{P \in \mathcal{P}} \ \mathcal{O}(P, u, \xi, D^\text{exp}) \quad \quad
     \text{subject to} \{u, \xi\} \text{ satisfying } \eqref{eq:gen_eqil},
\end{aligned} 
\end{equation}
where $\mathcal{P}$ defines a physical range of parameters and $\mathcal{O}(P, u, \xi. D)$ is the objective or loss function.  

We need to choose an objective function $\mathcal{O}$ that is both computationally efficient and one that attains a minimum when the displacement and forces computed using the solutions $\{ u, \xi\}$ to (\ref{eq:gen_eqil}) are equal to the measured values $D^\text{exp}$.  A somewhat subtle point in internal variable theories like (\ref{eq:gen_cons_law}) is that the internal variable is only defined pointwise, and thus does not have a mathematical meaning (specifically a trace) on the boundary.  However, the traction is mathematically defined through the interior stress distribution.  Thus, we have to interpret the measurement of boundary forces accordingly.  In this work we use a finite element discretization, and the governing equations are used in their weak form.  Specifically, the displacements are imposed on the nodes, but the strains, stress and internal variables are defined on the quadrature points and inherit their meaning by integration.   Therefore, it is natural to write the objective as a volume integral even though the force measurements are made on the boundary.  So, we take 
\begin{equation} \label{eq:obj}
    \mathcal{O}(P, u, \xi, D^\text{exp}) := \int_0^T \int_\Omega o(P, u, \xi, D^\text{exp}) \ d\Omega \ dt.
\end{equation}
for some $o$ that is concentrated near the boundary.  We discuss specific choices of $o$ in Section~\ref{sec:dem}.

%
%

\subsubsection{Adjoint method for sensitivity}  \label{sec:adj}
A gradient based optimization approach requires that we compute the sensitivities, that is, the total derivative of the objective with respect to the parameter set $P$ while enforcing the PDE constraint of the governing equations.  We may compute this by applying standard chain rule. However, this would require expressions for $\dv{u}{P}$ and $\dv{\xi}{P}$, which are the sensitivity of the solutions of (\ref{eq:gen_eqil}) to changes in parameters.  These are difficult to compute.  Therefore, we use the adjoint method \cite{optimization,Plessix2006} to circumvent the difficulty.   We rewrite the objective by adding a term that is zero according to the weak form of the evolution equation, 
\begin{equation}
\begin{aligned}
    \mathcal{O} = &\int_0^T \int_\Omega o(P, u, \xi, D^\text{exp}) \ d\Omega \ dt \\
    &+ \int_0^T \left \{ \int_\Omega \left[ - \rho \ddot{u} \cdot v - \sigma \cdot \nabla v + b \cdot v - \phi (\dot{\xi} - F) \right] \ d\Omega + \int_{\partial_f \Omega} f \cdot v \ dS \right \} \ dt,
\end{aligned}
\end{equation}
where $v$ and $\phi$ are test functions associated with momentum balance and internal variable evolution. The idea of the adjoint method is to choose the test functions $v$ and $\phi$ such that there is no need to explicitly compute the sensitivity of the solutions of (\ref{eq:gen_eqil}) to changes in parameters ($\dv{u}{P}$ and $\dv{\xi}{P}$).  Differentiating this rewritten objective with respect to $P$ gives
\begin{equation}
\begin{aligned}
    \dv{\mathcal{O}}{P} = \int_0^T \int_\Omega &\Biggr[  \pdv{o}{P} + \pdv{o}{u} \dv{u}{P} + \pdv{o}{\xi} \dv{\xi}{P} - \rho \dv{\ddot{u}}{P} \cdot v - \left( \pdv{\sigma}{P} + \pdv{\sigma}{\nabla u} \cdot  \nabla \dv{u}{P} + \pdv{\sigma}{\xi} \dv{\xi}{P} \right) \cdot \nabla v \\
    &  - \phi \left( \dv{\dot{\xi}}{P} - \pdv{F}{P} - \pdv{F}{\nabla u} \cdot \nabla \dv{u}{P} - \pdv{F}{\xi} \dv{\xi}{P} \right) \Biggr ]\ d\Omega \ dt.
\end{aligned}
\end{equation}
We group terms to obtain
\begin{equation}
    \begin{aligned}
    \dv{\mathcal{O}}{P} = \int_0^T \int_\Omega &\Biggr[  \left( \pdv{o}{P} - \pdv{\sigma}{P} \cdot \nabla v   + \phi \pdv{F}{P} \right) - \rho v \cdot \dv{\ddot{u}}{P} - \left( \nabla v \cdot \pdv{\sigma}{\nabla u}  - \phi \pdv{F}{\nabla u} \right) \cdot \left(\nabla \left(\dv{u}{P}\right)  \right) \\
    &  + \pdv{o}{u} \dv{u}{P} - \phi  \dv{\dot{\xi}}{P}  + \left( \pdv{o}{\xi} - \nabla v \cdot \pdv{\sigma}{\xi} +  \pdv{F}{\xi} \phi \right) \dv{\xi}{P} \Biggr ]\ d\Omega \ dt.
\end{aligned}
\end{equation}
The terms that include $\dv{\ddot{u}}{P}$ and $\dv{\dot{\xi}}{P}$ are integrated by parts temporally, while the term that includes $\nabla \left( \dv{u}{P} \right)$ is treated with divergence theorem in space. This yields
\begin{equation}
    \begin{aligned}
    \dv{\mathcal{O}}{P} = \int_0^T \int_\Omega &\Biggr[  \left( \pdv{o}{P} - \pdv{\sigma}{P} \cdot \nabla v   + \phi \pdv{F}{P} \right) + \left[ -\rho \ddot{v} + \pdv{o}{u} + \nabla \cdot \left( \nabla v \cdot \pdv{\sigma}{\nabla u}  - \phi \pdv{F}{\nabla u} \right) \right] \cdot \dv{u}{P}   \\
    &   + \left( \dot{\phi} +  \pdv{o}{\xi} - \nabla v \cdot \pdv{\sigma}{\xi} +  \pdv{F}{\xi} \phi \right) \dv{\xi}{P} \Biggr ]\ d\Omega \ dt \\
    &\hspace*{-4em} - \int_{0}^T \int_{\partial \Omega}   \left( \nabla v \cdot \pdv{\sigma}{\nabla u}  - \phi \pdv{F}{\nabla u} \right)n \cdot \dv{u}{P}  \ dS \ dt  \\
    &\hspace{-4em} + \left[ \int_{\Omega} \left( - \rho v \cdot \dv{\dot{u}}{P} +  \rho \dot{v} \cdot \dv{u}{P} - \phi \dv{\xi}{P} \right) \ d\Omega \right]_0^T .
\end{aligned}
\end{equation}
We eliminate the dependence on $\dv{u}{P}$ and $\dv{\xi}{P}$ by choosing a particular $v$ and $\phi$ such that the respective terms multiplying them become zero. As $\left .\dv{\dot{u}}{P} \right |_{t = 0} = \left .\dv{u}{P} \right |_{t = 0}$ = 0, and $\left .\dv{\xi}{P} \right |_{t = 0} = 0$, we eliminate the contributions from boundary terms by choosing $v|_{t = T} = \dot{v}_{t = T} = 0$ and $\phi |_{t = T} = 0$. 

We conclude that the expression for the sensitivity is
\begin{equation} \label{eq:gen_sens}
    \dv{\mathcal{O}}{P} = \int_0^T \int_\Omega \Biggr[  \pdv{o}{P} - \pdv{\sigma}{P} \cdot \nabla v   + \phi \pdv{F}{P}  \Biggr] \ d\Omega \ dt
\end{equation}
where the adjoint variables $v$ and $\phi$ satisfy the evolution
\begin{equation} \label{eq:gen_adj}
\begin{aligned}
    &\rho \ddot{v} - \nabla \cdot \left(\nabla v  \cdot \pdv{\sigma}{\nabla u} - \phi \pdv{F}{\nabla u} \right) =  \pdv{o}{u} &&\text{ on } \Omega, \\
    &\dot{\phi} = \nabla v \cdot \pdv{\sigma}{\xi} - \pdv{F}{\xi} \phi - \pdv{o}{\xi}  && \text{ on } \Omega, \\
    &\left(\nabla v  \cdot \pdv{\sigma}{\nabla u} - \phi \pdv{F}{\nabla u} \right) n = 0 &&\text{ on } \partial_f \Omega,    \\
    &v = 0 &&\text{ on } \partial_d \Omega,    \\
    &v |_{t = T} = \dot{v} |_{t = T} = 0, \ \phi |_{t = T} = 0. &&
\end{aligned}
\end{equation}
We note that the conditions for the adjoint variables are given at the end time $t = T$. Thus, the adjoint system should be solved backwards in time starting from $t = T$ and moving to $t = 0$. 
\vspace{\baselineskip}

\begin{figure}
    \centering
    \includegraphics[width=0.9\linewidth]{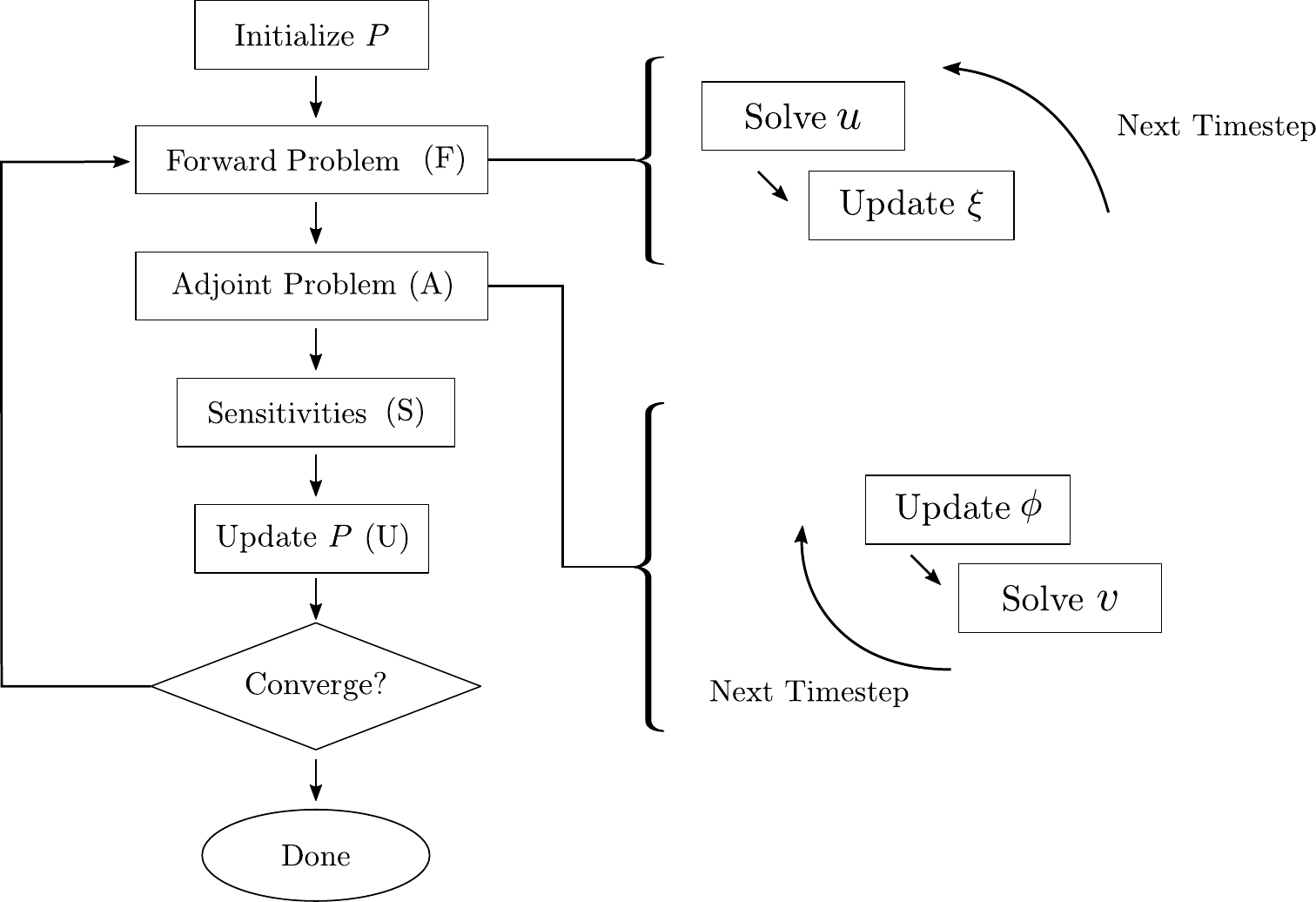}
    \caption{Schematic representation of the iterative algorithm to obtain to find material parameters. The forward and adjoint problem are given by (\ref{eq:gen_eqil}) and (\ref{eq:gen_adj}) respectively. The sensitivity  is given by  (\ref{eq:gen_sens}).}
    \label{fig:schematic_full_flow}
\end{figure}

Putting all of this together, we use the following procedure to solve the optimization problem (\ref{eq:gen_opt}).  We start with an initial value for the parameter set $P$.  We then solve the forward problem~\eqref{eq:gen_eqil} for $u(t),\ \xi(t)$ followed by the adjoint problem~\eqref{eq:gen_adj} for $v(t),\ \phi(t)$. These are used in~\eqref{eq:gen_sens} to compute the sensitivities. The sensitivities are used to update the parameters and the process is repeated until convergence. The schematic for the algorithm is shown in Figure \ref{fig:schematic_full_flow}.

\subsubsection{Scaling}
\label{sec:scaling}

We now discuss how the numerical cost of the algorithm shown in Figure~\ref{fig:schematic_full_flow} scales respect to the number of parameters.  The forward problem ((\ref{eq:gen_eqil}) labeled (F) in the figure) is largely independent of the number of parameters, as is the adjoint problem ((\ref{eq:gen_adj}) labeled (A)).  The calculation of the sensitivity ((\ref{eq:gen_sens}) labeled (S)) and the parameter update (labeled (U)) scale linearly, but do not involve the solution of any equations.  Thus the overall cost of each iteration is 
\begin{equation}
\text{Cost} (N_P) = A + B N_P, \quad \quad \text{where} \ A >> B
\end{equation}
where $N_P$ is the number of parameters.  In other words, the cost is linear with a small coefficient.  This is in contrast with gradient-free approaches that scale as $A \alpha^{N_P}$ for some $\alpha$ since $N_P$ is the dimension of the search space.


\subsection{J2 Elasto-viscoplasticity}
\label{sec:J2_plastic}

\subsubsection{Governing equations} \label{sec:J2f}

We now specialize to a specific example of a J2 plastic material with isotropic power-law hardening and rate dependence in the small strain setting~\cite{Ortiz1999,lubliner2008plasticity}.  We denote the displacement as $u$ and the strain as $\varepsilon = (\nabla u + \nabla u^T)/2$.  The plastic strain is $\varepsilon^p$ and the accumulated plastic strain is
\begin{equation}
    \dot{q} = \sqrt{\frac{2}{3} \dot{\varepsilon^p} \cdot \dot{\varepsilon^p}}.
\end{equation}
The body is linearly elastic before yield, with the stress and stored elastic energy density are
\begin{equation}
\sigma = \mathbb{C} \varepsilon^e \quad \text{and} \quad W^e = \frac{1}{2} \varepsilon^e \cdot {\mathbb C} \varepsilon^e
\end{equation}
respectively where $\mathbb{C}$ is the elastic modulus and  $\varepsilon^e = \varepsilon -\varepsilon^p$ is the elastic strain.  We assume a von Mises yield criterion with yield strength $\sigma_y$, power law isotropic hardening with potential 
\begin{equation}
    W^p(q) = \sigma_y\left[ q + \frac{n \varepsilon^p_0}{n + 1}\left( \frac{q}{\varepsilon^p_0}\right)^{(n+1)/n}\right],
\end{equation}
with $\varepsilon^p_0$ the reference plastic strain and $n$  the hardening exponent, and isotropic power law rate hardening with dissipation potential 
\begin{equation}
    \psi (\dot{q}) = \begin{cases} 
     g^*(\dot{q}) := \frac{ m \sigma_y \dot{\varepsilon}^p_0}{m + 1}\left( \frac{\dot{q}}{\dot{\varepsilon}^p_0}\right)^{(m + 1)/m} & \ \dot{q} \geq 0, \\
    \infty & \ \dot{q} < 0
    \end{cases},
\end{equation}
where $\dot{\varepsilon}^p_0$ is the reference plastic strain rate, and $m$ is the rate-sensitivity exponent. 

The governing equations for the fields  $\{u, q, \varepsilon^p \}$ are
\begin{equation} \label{eq:dyn_evo}
    \begin{aligned}
    \rho \ddot u &= \nabla \cdot {\mathbb C} \varepsilon^e  \qquad  &&  \text{ on } \Omega, \\
    0 &\in \sigma_M - \pdv{W^p}{q} - \partial \psi (\dot{q}) \quad && \text{ on } \Omega, \\
    \dot{\varepsilon}^p &= \dot{q} M  && \text{ on } \Omega, \\
    u &= u_0 &&  \text{ on } \partial_u \Omega\\
    f & = {\mathbb C} \varepsilon^e \cdot \hat n  &&  \text{ on } \partial_f \Omega\\
    & \hspace{-1em} u|_{t = 0} = \dot{u}|_{t = 0} = 0, \ q|_{t = 0} = 0,\ \varepsilon^p|_{t = 0} = 0, &&
    \end{aligned}
\end{equation}
where $u_0$ is the applied displacement on $\partial_u \Omega$, $f$ is the applied (dead) traction on  $\partial_u \Omega$, $\sigma_M =\sqrt{ \frac{2}{3} s \cdot s }$ is the Mises stress, $s := \mathbb{C} \varepsilon^e - (1/N) \trace(\mathbb{C} \varepsilon^e)$ is the stress deviator, and $M := (1/\sigma_M) s $ is the flow direction.  We have assumed quiescent initial conditions.

The governing equations for quasistatic evolution can be obtained by ignoring the inertial term in the momentum balance, and the initial conditions for the displacement.

\subsubsection{Adjoint method for sensitivities}
\label{sec:opt_for_viscoplastic_param}

We seek to find the material parameters by solving the optimization problem (\ref{eq:gen_opt}) with (\ref{eq:dyn_evo}) replacing (\ref{eq:gen_eqil}) as the constraint for a suitable objective $\mathcal O$.  We follow Section \ref{sec:adj} to find the 
sensitivity using the adjoint method; the details are provided in Appendix \ref{app:adj}.  The sensitivity is
\begin{equation} \label{eq:sensitivities}
		\dv{\mathcal{O}}{P} = \int_{0}^T  \int_{\Omega} \bigg [   \gamma \dot{q} \left( - \pdv{^2W^p}{q \partial P} - \pdv{^2 {g}^*}{\dot{q} \partial P}\right) \bigg ] \, d\Omega \, dt,
\end{equation}
where the adjoint variables $v$, $\gamma$, and $\zeta$, being dual to the forward variables $u$, $q$, and $\varepsilon^p$, satisfy the evolution relations
\begin{equation} \label{eq:vpadj}
    \begin{aligned}
    & \rho \ddot{v} - \nabla \cdot  \left(  \mathbb{C} \nabla v +  \gamma \dot{q} \pdv{\sigma_M}{\varepsilon} - \dot{q} \zeta \cdot \pdv{M}{\varepsilon} \right) =   - \pdv{o}{u} \cdot   && \text{ on } \Omega \\
    & \dv{}{t} \left[ \gamma \left( \sigma_M - \pdv{W^p}{q} - \pdv{g^*}{\dot{q}}\right) - \gamma \dot{q} \pdv{^2 g^*}{\dot{q}^2} - \zeta \cdot M \right]  = - \gamma \dot{q}  \pdv{^2W^p}{q^2} && \text{ on } \Omega \\
    & \dv{\zeta}{t} =  \nabla \xi \cdot \pdv{^2 W^e}{\varepsilon \partial \varepsilon^p} + \gamma \dot{q} \pdv{\sigma_M}{\varepsilon^p} - \dot{q} \zeta \cdot \pdv{M}{\varepsilon^p} && \text{ on } \Omega \\
     & v|_{t = T} = \dot{v} |_{t = T} = 0, \ \gamma |_{t = T} = 0, \quad  \zeta |_{t = T} = 0.  
    \end{aligned}
\end{equation}
Note that these equations are solved backward in time as before. 

We obtain the equation in the quasistatic setting by ignoring the inertial ($\ddot{v}$) term  in the first equation, and the final conditions on $\dot{v}$ in the last.

\subsection{Numerical implementation} \label{sec:num}

We discretize the equations in space using standard $P = 1$ Largrange polynomial finite elements for the displacement $u$ and the corresponding adjoint variable $v$.  The plastic quantities $q$ and $\varepsilon^p$ are spatially discretized at quadrature points.  In dynamics, we adopt a mixed scheme for the temporal discretization.  For the governing equations (\ref{eq:dyn_evo}), we use an explicit central difference update for the displacement, but an implicit backwards Euler scheme for the plastic updates.  Similarly, for the adjoint equations (\ref{eq:vpadj}) we use an explicit central difference method for the adjoint displacement $v$, and implicit backwards Euler scheme for the plastic adjoints $\gamma, \zeta$.  In quasistatic evolution, we use an implicit backward Euler scheme for temporal discretization of all variables.  Further details are provided in Appendix \ref{ap:a1}.

We solve the optimization problem iteratively starting from an arbitrary initial guess.  We update the parameter set using the gradient-based Method of Moving Asymptotes optimization scheme~\cite{svanberg1987}. This technique for solving constrained optimization problems has been used extensively in the optimal design community to solve PDE-constrained design problems over both small~\cite{AkersonLiu2023} and large parameter sets~\cite{Pedersen2003,Akerson2022}. We allow the optimization to run for a set number of iterations, with this number chosen large enough such that it sufficiently converges. We implement the numerical method in the deal.ii Finite Element Library~\cite{Bangerth2007,Arndt2021}.

\section{Demonstration and validation}
\label{sec:dem}

We now demonstrate and validate the method described above using two examples with synthetic data.   The first example is a quasistatic compression test on a plane-strain specimen while the second is an extended dynamic impact test.  In both cases, we generate synthetic data using numerical simulations of a material model with properties similar to that of copper.   Specifically, we take the shear modulus to be $\mu = 46.7 $ GPa, a Poisson ratio of $\nu = 0.3656$, and the density to be $\rho = 8.93 \times 10^{3}\ \text{kg/}\text{m}^3$, with the other parameters $P= \{\sigma_y,\ \varepsilon_0^p,\ n,\ \dot{\varepsilon}_0^p,\ m \}$ as shown in the the first row ($P^\text{gen}$) of Table~\ref{tab:params_full_table}\footnote{The stress scales as the shear modulus $\mu$ and therefore we use it to normalize all our calculations.}.  We use these parameters, and the forward problem described in Section~\ref{sec:J2f} above, to simulate the respective experiments to generate the data $D^\text{exp}$.  We then use this data to obtain the material parameters using the indirect inversion method described in Section~\ref{sec:J2_plastic}.   We initialize these calculations with an initial guess $P^\text{init}$ that are significantly different from the ground truth $P^\text{gen}$.  The $P^\text{gen}, P^\text{init}$ and recovered parameters for both experiments are summarized in Table \ref{tab:params_full_table}.

\begin{table}[t]
\begin{center}
\begin{tabular}{ |p{1.5cm} ||p{1.5cm}|p{1.5cm}|p{1.4cm}|p{1.5cm}|p{1.4cm}|p{1.9cm}|p{1.5cm}|}
\hline
\multicolumn{8}{|c|}{\textbf{Synthetic and Converged Elasto-Viscoplastic Material Parameters}} \\
\hline
\ & $\sigma_y/\mu$ & $\varepsilon^p_0$ & $n$  & $\dot{\varepsilon}^p_0$  ($/s$) & $m$ & $\tilde{\mathcal{O}}$& $\mathcal{O}_\text{ind}$\\
\hline
$P^\text{gen}$ & 0.001935 &  0.02245&  3.23 & 5.00$\times10^5$ & 2.00 & -- & -- \\
$P^\text{init}$ & 0.005000 & 0.05000 & 5.00 & 3.00$\times10^5$& 5.00 & -- & -- \\
$P^\text{QS}$ & 0.002127 & 0.03932  & 3.05 & 4.75$\times10^5$& 1.88 & 5.59$\times 10^{-9}$ (2.74$\times 10^{-4}$) & 0.0069 (1.452) \\
$P^\text{DC}$ & 0.001734 & 0.02859 & 2.87 & 4.05$\times10^5$& 4.50 & 1.44$\times10^{-4}$ (1.81$\times10^{-1}$) & 0.021 (1.452) \\
\hline
\end{tabular}
\end{center}
\caption{Summary of results.  The parameters used to generate the synthetic data, the initial guess and ones obtained through the proposed method in the quasistatic tension (QS) and dynamic compression(DC) experiments.  The final objectives are reported in the table with the initial objectives in the parenthesis.  }
\label{tab:params_full_table}
\end{table}

\subsection{Quasistatic compression test of a plane strain specimen} \label{sec:syn_qs}
\begin{figure}[t]
\centering
\includegraphics[width=0.9\textwidth]{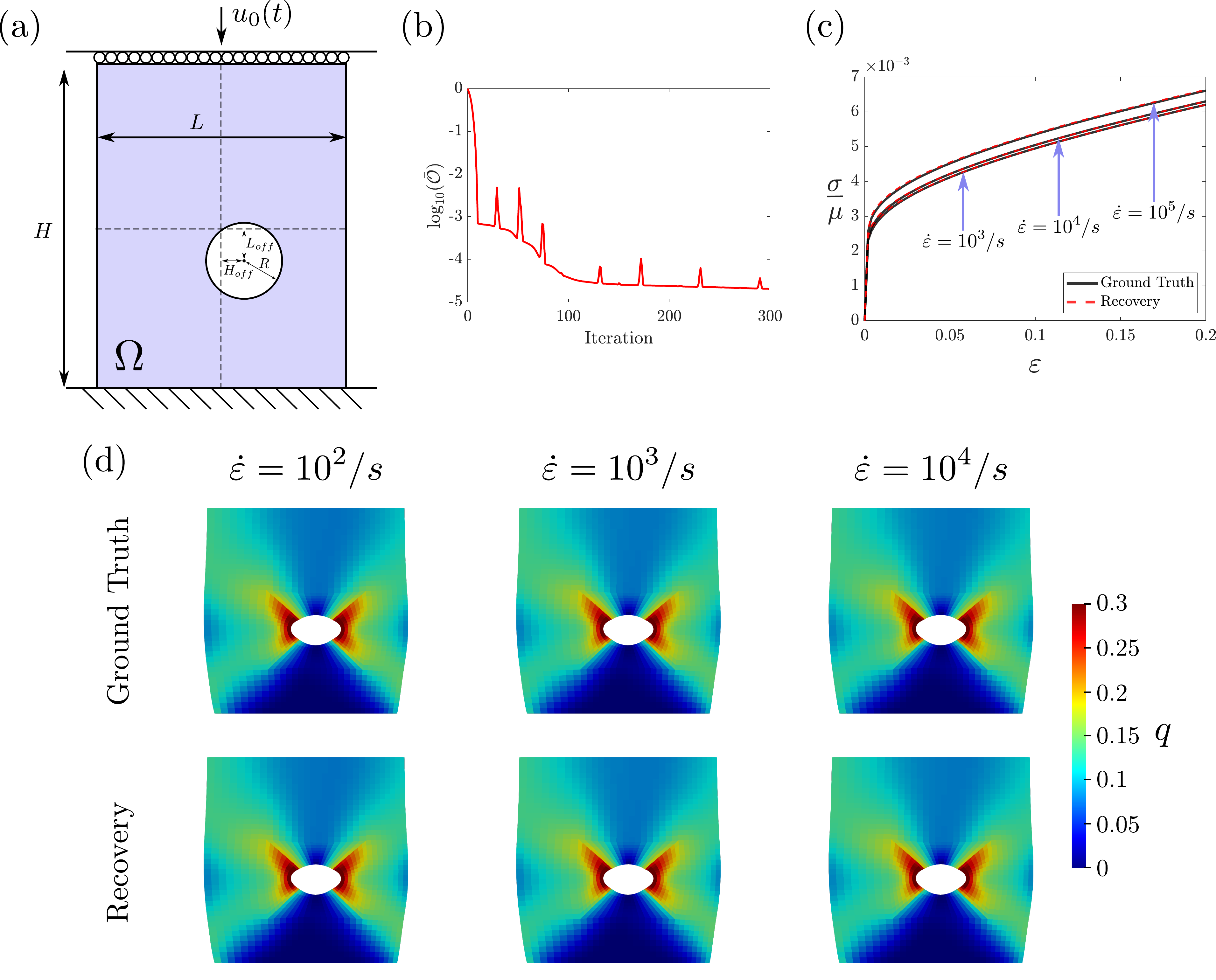}
\caption{ 
Results for the quasistatic compression of a thick plate with offset hole. (a) Geometry and boundary conditions. (b) Normalized objective from \ref{eq:oqs} vs iteration number. (c) Results of an independent uniaxial stress-strain test. (d) Deformed configurations at the final time-step for the ground truth data with the accumulated plasticity $q$. 
\label{fig:qs}}
\end{figure}

The first example is a thick plate with an off-center hole subjected to quasistatic compression shown in Figure \ref{fig:qs}(a). We consider a geometry with $L/H = 0.8$, $R/H = 0.1$, $L_\text{off}/H = 0.1$, and $H_\text{off}/H = 0.04$. We assume plane strain conditions. We impose uniform vertical displacement 
\begin{equation}
u_0(t) = \dot{\bar{\varepsilon}} H t,
\end{equation}
with a nominal or macroscopic strain rate $\dot{\bar{\varepsilon}}$ on the top boundary, $\mathcal{T}$, while the botton surface is held fixed.The experimental data $D^\text{exp}$ consists of the full-field displacement $u^\text{exp}(x,T) = u(x,T)$ at the final time $T$, and the reaction load history $f_R^\text{exp}(t) := \int_{\mathcal T} \mathbb{C} \varepsilon^e n \ d\Gamma$.   We emphasize that we only use the full-field displacement at the final time step\footnote{It is typical in such tests to use a high speed camera to capture multiple snapshots and thus have the displacement field over many snapshots.  We could extend our method to include multiple snapshots, but we chose not to do so to understand how much information we can obtain from a minimal amount of data since the inversion from DIC to displacement fields can also add computational cost.}.  We further assume that we have data from $n$ tests with varying rates.  We define our objective to be 
\begin{equation} \label{eq:oqs}
    \begin{aligned}
    \mathcal{O}(P) & := \sum_{i = 1}^n \left( \frac{\alpha_u }{L^4}\int_\Omega |u^i (x,T) - u^{\text{exp},i}(x,T)|^2 d \Omega +
    \frac{\alpha_f}{T_i \mu^2 L^2} \int_0^{T_i} \norm{ f_u^i(t) - f^{\text{exp},i}_R }^2 \ dt \right)\\
    & \quad \quad \text{where $\{u^i, q^i, \varepsilon^{p,i} \}$ satisfy the governing equation for each $i = 1, \dots, n$} 
    \end{aligned}
\end{equation}
where the superscript indexes each of the tests, and $\alpha_u$ and $\alpha_f$ are scaling factors that balance the weights of the two objective terms.

We consider a data-set from by three tests ($n=3$) on the same geometry and material, but differing in the loading rate. The loading rates span two orders of magnitude, with macroscopic strain rates of $10^2$, $10^3$, and $10^4$/s, reaching a final macroscopic strain of $\dot{\bar{\varepsilon}} = 0.1$ over $100$ time-steps. The data-set is generated  synthetically from forward simulations using the parameters  $P^\text{gen}$ in Table~\ref{tab:params_full_table}.  We then use the proposed approach with the objective (\ref{eq:oqs}). Full numerical details of the forward and adjoint problem  can be found in Appendix~\ref{ap:forward_qs} and Appendix~\ref{ap:adjoint_qs}

The results are shown in Figure \ref{fig:qs}(b-d) and the recovered parameters are listed in Table~\ref{tab:params_full_table}.  
Figure~\ref{fig:qs}(b) shows the change in the objective as we iterate.  It drops rapidly but then gradually stabilizes after a little over 100 iterations decreasing by a roughly a factor $10^5$.  The values of the recovered parameters after 300 iterations is listed as $P^\text{QS}$ in the Table~\ref{tab:params_full_table}.   We recover the normalized yield strength $\sigma_y/\mu$ to within 10\%, but the strain and rate hardening parameters differ significantly.  This is true despite the fact that our objective is extremely small, with an objective value on the order of $10^{-9}$. This means that the experiment with three tests at three strain rates are unable to distinguish between the two sets of parameters, the parameters $P^\text{gen}$ used to generate the data and the recovered parameters $P^\text{rec}$.  This is further demonstrated in Figure \ref{fig:qs}(d) which compares the accumulated plastic strain computed using the two sets of parameters, $P^\text{gen}$ and $P^\text{QS}$.

\paragraph{Independent objective}
We found that the parameters we recover $P^\text{QS}$ is different from those used to generate the data $P^\text{gen}$ even though the difference in the measured quantities (objective) are very small.  Therefore, we consider a {\it zero-shot test} where we evaluate our results against an independent objective that is not used in the inverse problem.  We simulate the material response with the two sets of parameters with an independent uniaxial tension test.  The resulting stress-strain curves are shown in Figure \ref{fig:qs}(c), and they agree well.  Thus, even the independent uniaxial stress tension test is also unable to distinguish between the two sets of parameters.

To assess this quantitatively, we define a {\it independent objective} to be the average relative root mean square error of the material response for tests performed at $n$ different strain rates:
\begin{align}
    \mathcal{O}_\text{ind} = \frac{1}{n} \sum_{i=1}^{n}  \left( \frac{\int \left( \sigma_{\text{rec}}(\varepsilon) - \sigma_{\text{gen}}(\varepsilon) \right) ^2 d\varepsilon }{  \int  \sigma_{\text{gen}}^2 \ d\varepsilon} \right)^{1/2} .
\end{align}
where $\sigma_{\text{gen}}$ is the uniaxial stress computed using $P^\text{gen}$ while $\sigma_{\text{rec}}$ is the uniaxial stress-strain curve generated using the recovered parameters.   The value of $ \mathcal{O}_\text{ind}$ Table \ref{tab:params_full_table} confirms the results in Figure \ref{fig:qs}(c) that the two sets of parameters can not be distinguished in independent uniaxial stress-strain curves.
\vspace{\baselineskip}

We conclude this section by studying the robustness of the method to the initial guess, objective and mesh size.

\paragraph{Sensitivity to initial guess}
We consider three tests with different and quite distinct initialization sets $\{P^{\text{init}}_1, P^{\text{init}}_2, P^{\text{init}}_3 \}$. These are shown, along with their corresponding recovered parameter sets in Table~\ref{tab:params_study_QS}. We observe that the method is insensitive to the initial parameter guess, yielding only slight variations in the recovered parameters with the objective remaining on the order of $10^{-9}$ for all of the recovered parameter values.

\begin{table}[t]
\begin{center}
\begin{tabular}{ |p{2.1cm} ||p{1.5cm}|p{1.5cm}|p{1.5cm}|p{1.5cm}|p{1.5cm}|p{2cm}|p{1.5cm}|}
\hline
\ & $\sigma_y/\mu$ & $\varepsilon^p_0$ & $n$  & $\dot{\varepsilon}^p_0$  ($/s$) & $m$ & $\mathcal{O}$ & $\mathcal{O}_\mathcal{\text{ind}}$\\
\hline
$P^\text{gen}$ & 0.001935 &  0.02245&  3.23 & 5.00$\times10^5$ & 2.00&&\\
\hline
\multicolumn{8}{|c|}{Sensitivity to initial guess}\\
\hline
$P^\text{init}_1$  & 0.005000 & 0.05000 & 5.00 & 3.00$\times10^5$& 5.00 & 2.74$\times10^{-4}$ & 1.452\\
$P^\text{rec}_1$ & 0.002127 & 0.03932  & 3.05 & 4.75$\times10^5$& 1.88 &$ 5.59\times10^{-9}$& 0.0069 \\
$P^\text{init}_2$ & 0.01 & 0.1 & 2.0 & 2.0$\times10^5$& 6.0 & 6.43$\times10^{-4}$ & 3.791 \\
$P^\text{rec}_2$ & 0.002064 & 0.03352 & 3.09 & 8.16$\times 10^5$  & 2.23 & 2.81$\times10^{-9}$& 0.0060 \\
$P^\text{init}_3$ & 0.0075 & 0.075 & 4.0 & 4.0$\times10^5$& 1.0 & 3.49$\times10^{-4}$ & 1.91 \\
$P^\text{rec}_3$  & 0.002142 & 0.04039 & 3.05 & 3.62$\times 10^5$ & 1.71 & 7.37$\times10^{-9}$ & 0.0128 \\
\hline
\multicolumn{8}{|c|}{Sensitivity to objective}\\
\hline
$P^\text{gen}$ & 0.001935 &  0.02245&  3.23 & 5.00$\times10^5$ & 2.00& --& --\\
$P^\text{init}_1$ & 0.005000 & 0.05000 & 5.00 & 3.00$\times10^5$& 5.00 & -- & 1.452 \\
$\alpha_f/\alpha_u = 0.1$  & 0.002081 & 0.03456 & 3.08 & 5.34$\times10^5$& 1.96 & $4.44 \times10^{-9}$& 0.0049  \\
$\alpha_f/\alpha_u = 1$  & 0.002127 & 0.03932  & 3.05 & 4.75$\times10^5$& 1.88 &$ 5.59\times10^{-9}$& 0.0069  \\
$\alpha_f/\alpha_u = 10$  & 0.002132 & 0.04009 & 3.04 & 4.98$\times10^5$& 1.87 & $2.44\times10^{-9}$ & 0.0048 \\
\hline
\multicolumn{8}{|c|}{Sensitivity to meshsize}\\
\hline
$P^\text{gen}$ & 0.001935 &  0.02245&  3.23 & 5.00$\times10^5$ & 2.00& --& --\\
$P^\text{init}$ & 0.005000 & 0.05000 & 5.00 & 3.00$\times10^5$& 5.00 & -- & 1.452 \\
$h = h_0$  & 0.002036 & 0.03233 & 3.08 & 6.81$\times10^5$& 2.26 & $9.94 \times10^{-10}$ & 0.0031 \\
$h = 0.5\ h_0$ & 0.002127 & 0.03932  & 3.05 & 4.75$\times10^5$& 1.88 &$ 5.59\times10^{-9}$& 0.0060  \\
$h = 0.25\ h_0$  & 0.002150 & 0.04410 & 3.00 & 4.56$\times10^5$& 1.95 & $7.50\times10^{-9}$ & 0.0089\\
$h = 0.125\ h_0$  & 0.002129 & 0.04410  & 3.00 & 4.47$\times10^5$& 2.18 &$ 8.44\times10^{-9}$& 0.0090 \\
\hline
\end{tabular}
\end{center}
\caption{Robustness of the proposed method for the quasistatic experiment.}
\label{tab:params_study_QS}
\end{table}

\paragraph{Sensitivity to objective}
The parameters $\alpha_u$ and $\alpha_f$ in the objective function represent the weights for the displacement data and the force data, respectively. Table~\ref{tab:params_study_QS} shows the recovered set of parameters for three different orders of $\alpha_f/\alpha_u$, with all of them showing similar performance.  

\paragraph{Sensitivity to mesh size}
To confirm the convergence of solution with respect to the mesh size, we perform the optimization on different mesh refinement levels. Table~\ref{tab:params_study_QS} shows the recovered parameters for four mesh resolutions ranging from 384 to 24,576 elements. As the recovered parameters only differ slightly between the four meshes, we conclude that the solution does not depend on mesh resolution.
\vspace{\baselineskip}

We conclude that the method accurately recovers elasto-viscoplastic material parameters from data obtained from the final snapshot and force-history of three quasistatic tests. While the values of the recovered and generate parameters differ slightly, 
these parameters are unable to distinguish either observations of the test, or the zero-shot independent test, and reflects the degeneracy of the constitutive model.

%
%
%
%

\subsection{Dynamic compression of thin specimens} \label{sec:syn_dyn}


\begin{figure}[t]
\centering
\includegraphics[width=6in]{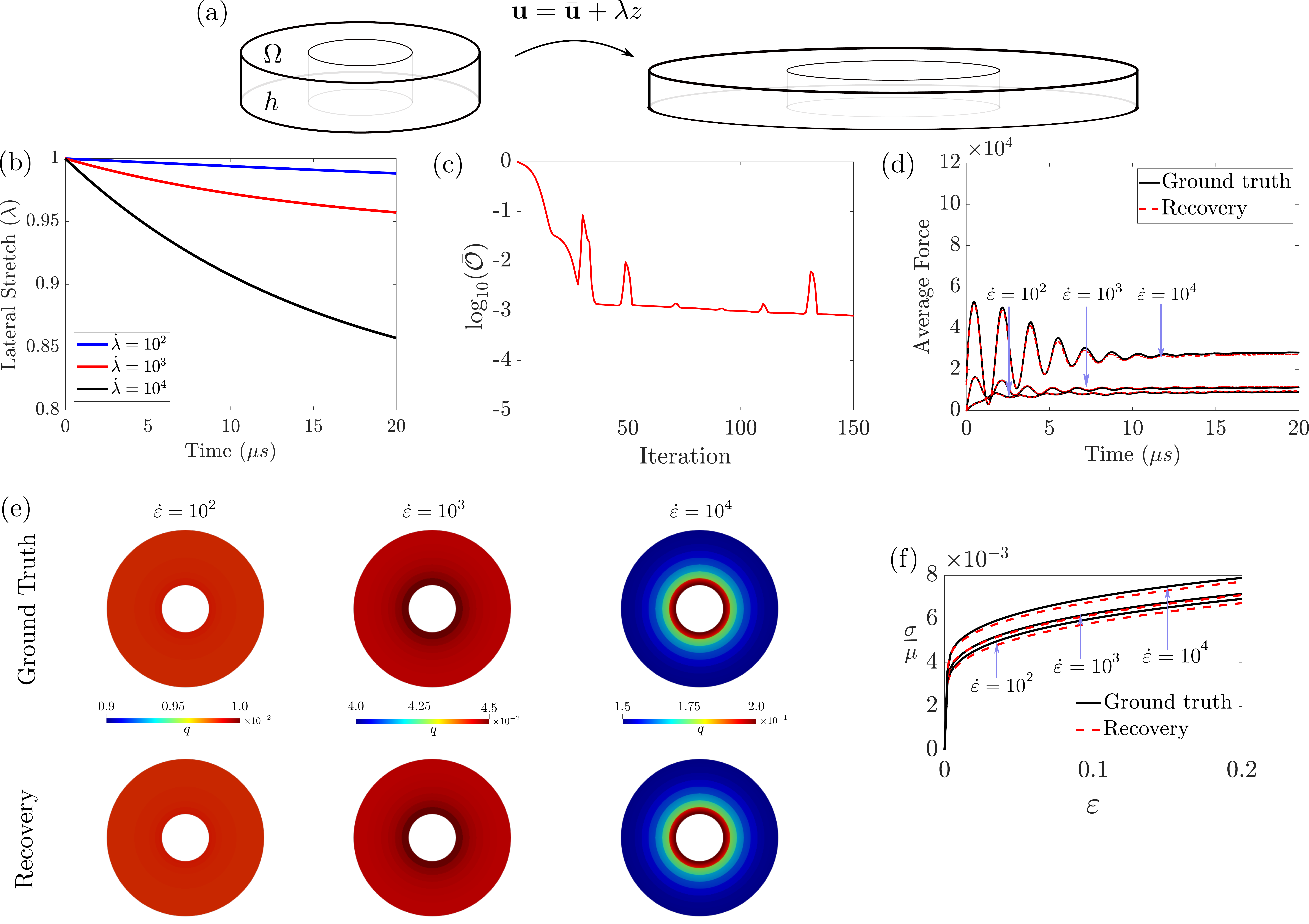}
\caption{ Results for the dynamic compression of circular annulus. (a) Geometry and deformation. (b) Applied stretch along the annulus thickness (c) Normalized objective from \ref{eq:opt_dyn} vs iteration number. (d) The net vertical force applied to the annulus over time. (e) Comparison of accumulated plasticity $q$.  (f) Results of an independent uniaxial stress-strain test. }
\label{fig:dy}
\end{figure}

The second example we consider is that of dynamic impact of a cylinder. We consider a thin cylindrical specimen that $\Omega := \overline{\Omega} \times (0, h)$ with cross-section $\overline{\Omega} \in \mathbb{R}^2$ and thickness of $h \in \mathbb{R}$.  Figure~\ref{fig:dy}(a) shows the case where $\overline{\Omega}$ is a circular annulus.  The cylinder is placed on an anvil and impacted on the top with a striker whose cross section is larger than that of the cylinder.  Ignoring an irrelevant rigid translation, this corresponds to the following imposed boundary conditions:
\begin{equation}
u_3 (X_1, X_2, 0, t) = 0, \quad u_3 (X_1, X_2, h, t) =  (\lambda(t) - 1) h
\end{equation}
where $\lambda(t)$ is the imposed nominal axial stretch.
We assume that the contact with the anvil and the striker is friction-free, and that the specimen is sufficiently small so that we may assume uniform axial strain,
\begin{equation}  \label{eq:u_assump}
    u (X_1, X_2, X_3, t) = \bar{u}(X_1, X_2, t) + (\lambda(t)-1) X_3 \ e_3,
\end{equation}
where $\bar{u} := \bar{\Omega} \mapsto \mathbb{R}^2$ is the in-plane displacement.   This enables us to reduce this to two space dimensions, see Appendix \ref{sec:annulus} for details.

Our experimental data $D^\text{exp}$ consists of the final in plane displacement $\bar{u}^\text{exp}(X_1,X_2,T)$ and the net axial force history $f_R^\text{exp}(t)$.  Notice that this is more data than is typically collected in a classical dynamic impact experiment, where only $f_R^\text{exp}$ is measured.  However, we may measure $\bar{u}^\text{exp}(X_1,X_2,T)$ by placing a speckle pattern on the face of the specimen, imaging it before and after the impact, and using digital image correlation.  We repeat the test $n$ times, and consider the following objective
\begin{equation} \label{eq:opt_dyn}
    \begin{aligned}
    \mathcal{O}(P) & := \sum_{i = 1}^n \left(\frac{\alpha_u }{L^4} \int_{\overline{\Omega}} |\bar{u}^i(X_1,X_2,T) - \bar{u}^{exp, i}(X_1,X_2,T)|^2 \ d \bar{\Omega} + \frac{\alpha_f}{T_i \mu^2 L^4}  \int_0^T |f_u^i - f^{exp,i}_R |^2 \ dT \right) \\
    & \quad \quad \text{where $\{\bar{u}^i, q^i, \varepsilon^{p,i} \}$ satisfy \eqref{eq:dyn_evo_red} for $i = 1, \dots, n$},
    \end{aligned}
\end{equation}
and $f^{exp,i}_u : \int_{\bar{\Omega}} (\mathbb{C} \bar \varepsilon^{e,i})_{33} \ dS$ is the net vertical force. $\alpha_u$ and $\alpha_f$ are weights associated with the displacement and force components of the objective. $L, \mu$ and $T_i$ are characteristic length, shear modulus, and characteristic time respectively. 

\begin{table}[t]
\begin{center}
\begin{tabular}{ |p{2.1cm} ||p{1.5cm}|p{1.5cm}|p{1.5cm}|p{1.5cm}|p{1.3cm}|p{1.8cm}|p{1.5cm}|}
\hline
\ & $\sigma_y/\mu$ & $\varepsilon^p_0$ & $n$  & $\dot{\varepsilon}^p_0$  ($/s$) & $m$ & $\mathcal{O}$ & $\mathcal{O}_\mathcal{\text{ind}}$\\
\hline
$P^\text{gen}$ & 0.001935 &  0.02245&  3.23 & 5.00$\times10^5$ & 2.00&&\\
\hline
\multicolumn{8}{|c|}{Sensitivity to initial guess}\\
\hline
$P^\text{init}_1$ & 0.005000 & 0.05000 & 5.00 & 3.00$\times10^5$& 5.00 & 1.81$\times10^{-1}$ & 1.452 \\
$P^\text{rec}_1$ & 0.001734 & 0.02859 & 2.87 & 4.05$\times10^5$& 4.50 &1.44$\times10^{-4}$&0.021\\
$P^\text{init}_2$ & 0.01 & 0.1 & 2.0 & 2.0$\times10^5$& 6.0 & 5.55$\times10^{-1}$ & 3.791 \\
$P^\text{rec}_2$ & 0.001793 & 0.06623 & 2.11 & 2.40$\times10^5$& 4.50 &8.57$\times10^{-4}$& 0.079\\
$P^\text{init}_3$ & 0.0075 & 0.075 & 4.0 & 4.0$\times10^5$& 1.0 & 2.12$\times10^{-1}$ & 1.91 \\
$P^\text{rec}_3$ & 0.0022 & 0.07812 & 2.70 & 6.02$\times10^5$& 1.09 & 2.93$\times10^{-4}$&0.090\\
\hline
\multicolumn{8}{|c|}{Sensitivity to objective}\\
\hline
$P^\text{gen}$ & 0.001935 &  0.02245&  3.23 & 5.00$\times10^5$ & 2.00 & -- & -- \\
$P^\text{init}_1$ & 0.005000 & 0.05000 & 5.00 & 3.00$\times10^5$& 5.00 & 1.81$\times10^{-1}$ & 1.452 \\
$\alpha_f/\alpha_u = 0.1$  & 0.001726 & 0.03583 & 2.71 & 3.46$\times10^5$& 4.58 & $1.47\times10^{-4}$& 0.050\\
$\alpha_f/\alpha_u = 1$  & 0.001734 & 0.02859 & 2.87 & 4.05$\times10^5$& 4.50 &1.44$\times10^{-4}$&0.021\\
$\alpha_f/\alpha_u = 10$  & 0.001778 & 0.03929 & 2.69 & 3.51$\times10^5$& 4.81 & $0.87\times10^{-4}$ & 0.043\\
\hline
\multicolumn{8}{|c|}{Sensitivity to geometry}\\
\hline
$P^\text{gen}$ & 0.001935 &  0.02245&  3.23 & 5.00$\times10^5$ & 2.00 & -- & -- \\
$P^\text{init}_1$ & 0.005000 & 0.05000 & 5.00 & 3.00$\times10^5$& 5.00 & 1.81$\times10^{-1}$ & 1.452 \\
Circular & 0.001734 & 0.02859 & 2.87 & 4.05$\times10^5$& 4.50 &1.44$\times10^{-4}$&0.021\\
Ellipse & 0.001921 & 0.04590 & 2.55 & 3.10$\times10^5$& 3.14 & $1.81\times10^{-5}$& 0.020 \\
Flower hole & 0.001731 & 0.03506 & 2.73 & 3.71$\times10^5$& 4.26 & $1.32\times 10^{-4}$& 0.021\\
Flower & 0.001800 & 0.02768 & 2.93 & 3.93$\times10^5$& 3.04 & $8.49\times 10^{-5}$& 0.023\\
\hline
\end{tabular}
\end{center}
\caption{Robustness of the proposed method for the dynamic compression experiment.}
\label{tab:params_study_Dyna}
\end{table}

\begin{figure}[t]
\centering
\includegraphics[width=2in]{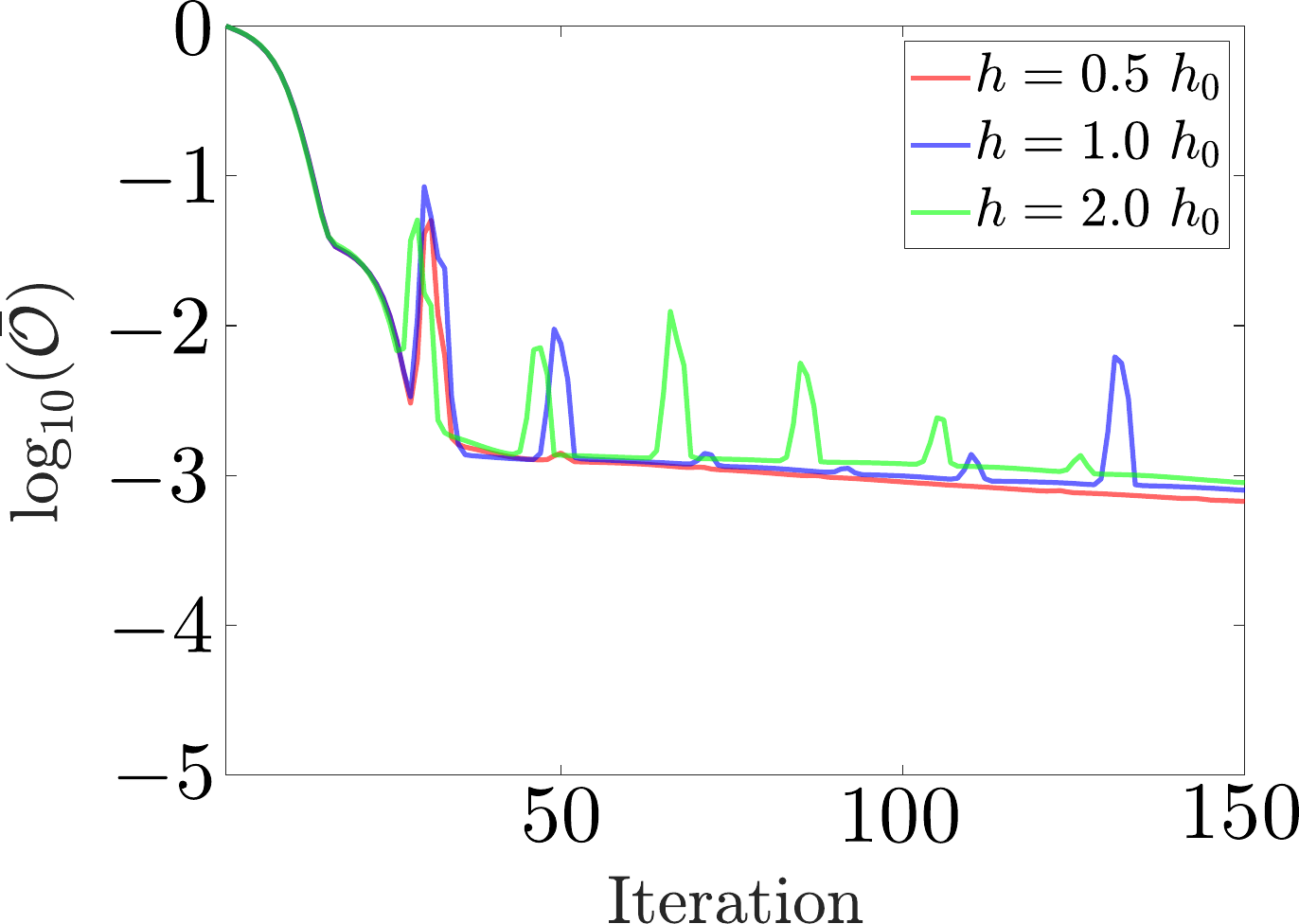}
\caption{Objective curves with respect to iterations for different mesh sizes. }\label{fig:mesh_conv_dy}
\end{figure}

We apply our method to a specimen whose cross-section is a circular annulus shown in Figure~\ref{fig:dy}(a).  We  generate synthetic data for $n=3$ tests imposed thickness strains shown in Figure~\ref{fig:dy}(b) with the initial strain rate in the range $\dot{\varepsilon} = 10^2 - 10^4$ s$^{-1}$ with the parameters $P^\text{gen}$.  We initialize our iterative optimization approach with the parameters $P^\text{init}$. $\alpha_u$ and $\alpha_f$ are chosen such that the contribution of both parts of the objective are approximately equal, $\alpha_f/\alpha_u = 1$. Figure~\ref{fig:dy}(c) shows the evolution of the objective with iteration.  The objective decreases rapidly at first and then stabilizes at a value with a factor  greater than 10$^3$ than the initial value.   The resulting parameters are shown in Table~\ref{tab:params_full_table} as $P^\text{DC}$.  As in the previous quasistatic compression test, there is good agreement with the yield strength, but not for the hardening parameters.  Despite the fact that the objective is very small, the experimental observations agree well, as shown in Figures~\ref{fig:dy}(d,e).  Figure~\ref{fig:dy}(d) compares the experimental reaction force with the one simulated with the recovered parameters, and we observe excellent agreement.  Figure~\ref{fig:dy}(e) compares the accumulated plastic strain computed with the original parameters $P^\text{gen}$ with those computed with the recovered parameters $P^\text{DC}$. Again, we see good agreement.  Finally, we conduct the zero-shot test of comparing the response in uniaxial tensile tests.  Figure~\ref{fig:dy}(f) shows the rsults of the independent stress-strain test, and we again see very good agreement.

Finally, we demonstrate the robustness of our method.
\paragraph{Sensitivity to initial guess} 
We consider three tests with different and quite distinct initialization sets $\{P^{\text{init}}_1, P^{\text{init}}_2, P^{\text{init}}_3 \}$. These are shown, along with their corresponding recovered parameter sets in Table~\ref{tab:params_study_Dyna}. We observe that the method is insensitive to the initial parameter guess, yielding only slight variations in the recovered parameters with the objective remaining on the order of $10^{-9}$ for all of the recovered parameter values.

\paragraph{Sensitivity to objective}
The parameters $\alpha_u$ and $\alpha_f$ in the objective function represent the weights for the displacement data and the force data respectively. Table~\ref{tab:params_study_Dyna} shows the recovered set of parameters for three different orders of $\alpha_f/\alpha_u$, with all of them showing similar performance.

\paragraph{Sensitivity to mesh size}
To confirm the convergence of solution with respect to the mesh size, we perform the optimization for three different mesh sizes. Figure~\ref{fig:mesh_conv_dy} shows the objective versus optimization iterations for different mesh sizes. Since the objective remains the same for all mesh sizes, we conclude that it is safe to assume the solution does not depend on the size of the mesh.  

\begin{figure}[t]
\centering
\includegraphics[width=6in]{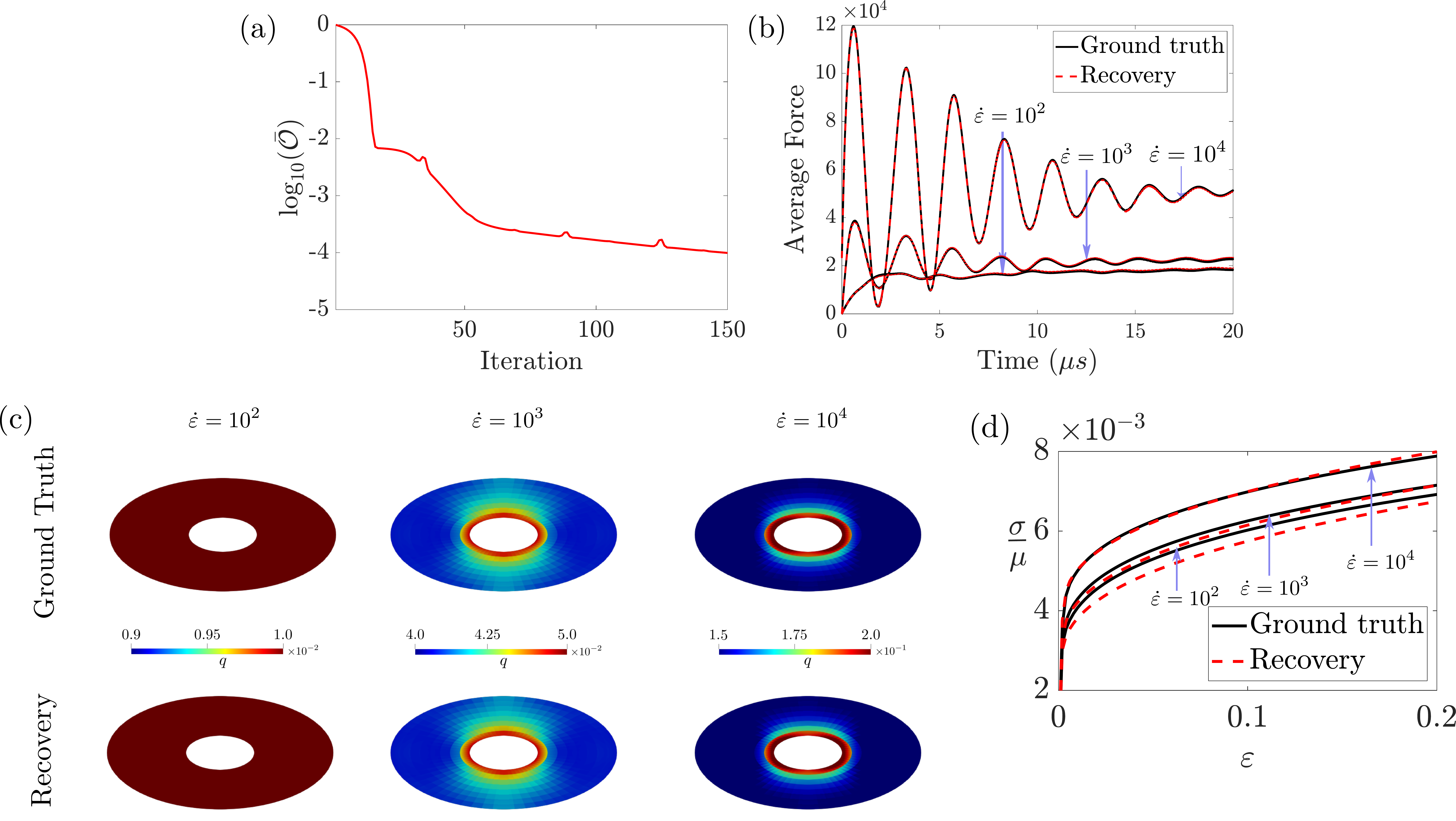}
\caption{ Results for the dynamic compression of elliptic annulus  \label{fig:dyn_ellip}. (a) Normalized objective from \ref{eq:opt_dyn} vs iteration number. (b) The net vertical force applied to the annulus over time. (c) Comparison of accumulated plasticity $q$.  (d) Results of an independent uniaxial stress-strain test. }
\end{figure}

\begin{figure}[t]
\centering
\includegraphics[width=6in]{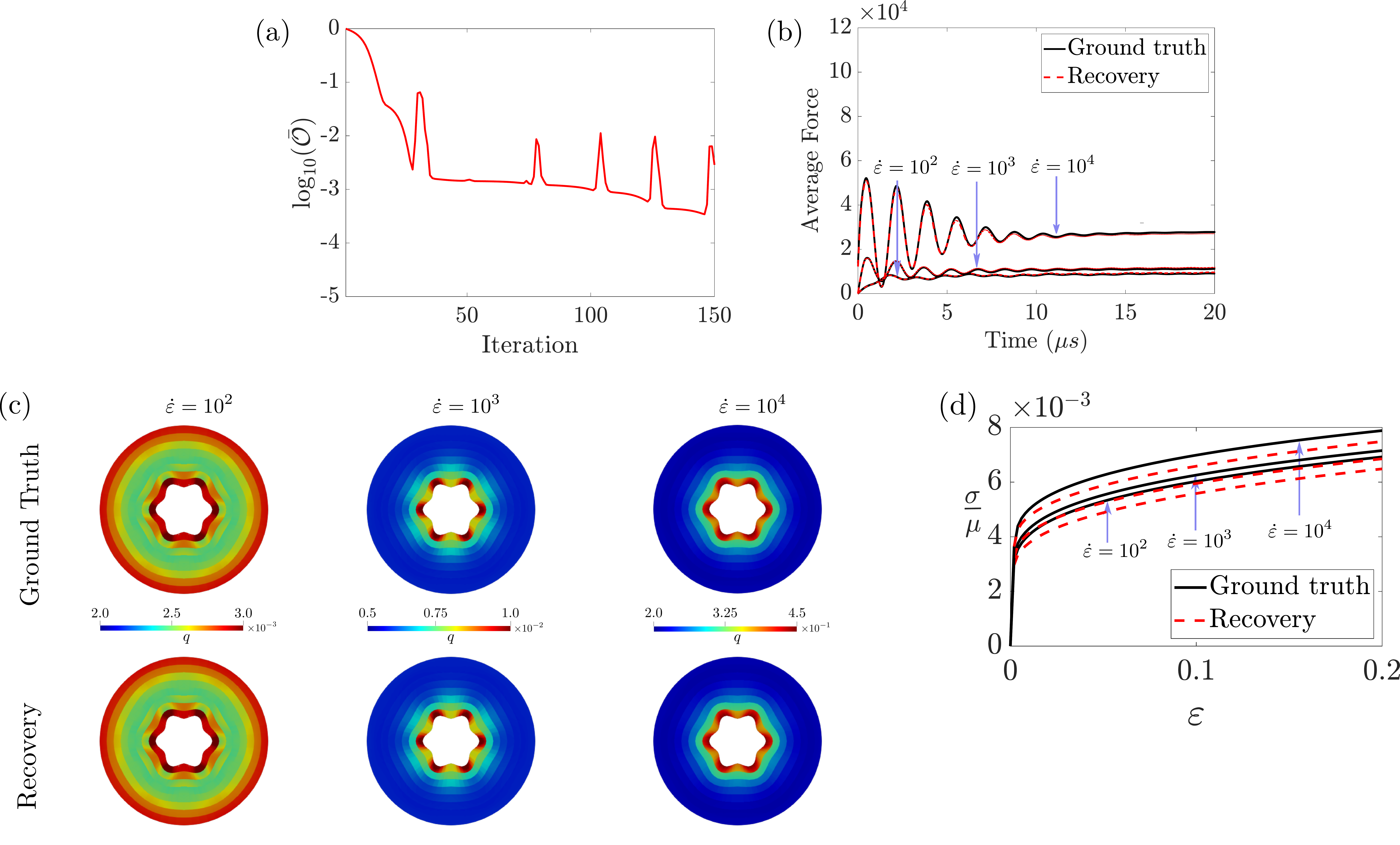}
\caption{ 
Results for the dynamic compression of  flower-shaped specimen  \label{fig:dyn_flr}. (a) Normalized objective from \ref{eq:opt_dyn} vs iteration number. (b) The net vertical force applied to the annulus over time. (c) Comparison of accumulated plasticity $q$.  (d) Results of an independent uniaxial stress-strain test. }
\end{figure}

\begin{figure}[t]
\centering
\includegraphics[width=6in]{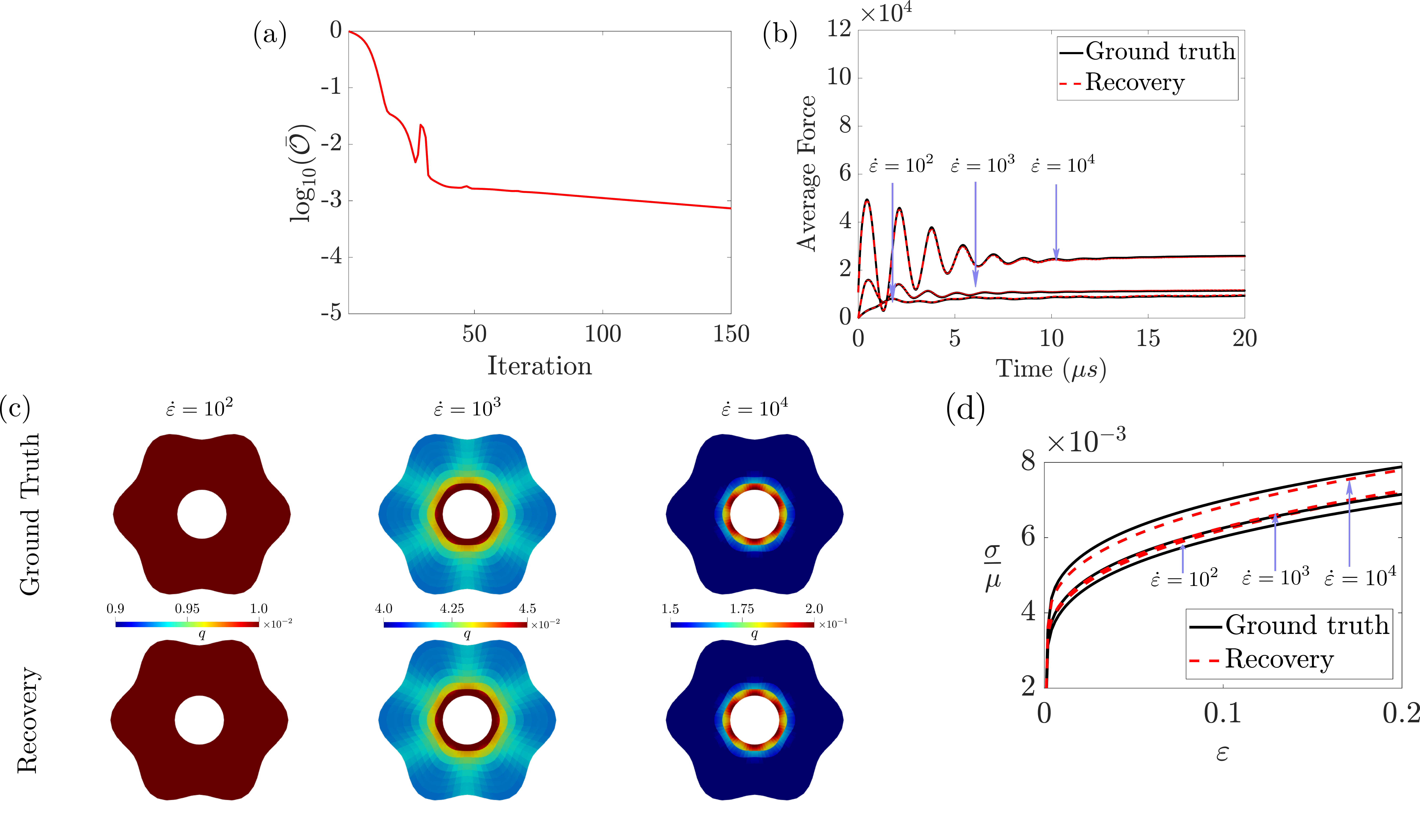}
\caption{ 
Results for the dynamic compression of disc with flower-shaped hole  \label{fig:dyn_invflr}. (a) Normalized objective from \ref{eq:opt_dyn} vs iteration number. (b) The net vertical force applied to the annulus over time. (c) Comparison of accumulated plasticity $q$.  (d) Results of an independent uniaxial stress-strain test. }
\end{figure}

\paragraph{Sensitivity to configuration}
The geometry of the specimen can significantly impact the inversion methodology. Certain geometries can induce a wide range of strains during compression, providing datasets with richer information about the material properties. To investigate this, we performed the inversion on different cross-sectional geometries, specifically circular, elliptical, and flower-shaped configurations. The results for the elliptic annulus, disc with flower-shaped hole and flower-shaped specimen are shown in Figures~\ref{fig:dyn_ellip}, \ref{fig:dyn_flr} and \ref{fig:dyn_invflr}, respectively. The recovered parameters for these configurations are compared in Table~\ref{tab:params_study_Dyna}. Due to the geometry of the ellipse configuration, the average forces experienced is much higher than the other two configurations, leading to more efficient recovery of parameters and a lower overall objective value. Furthermore, the lack of angular symmetry of the strain field in ellipse provides a richer dataset resulting in a more efficient inversion. 

\section{Conclusion}
\label{sec:conc}

Material characterization through a constitutive model is necessary to close the balance laws and allow for continuum modeling. Traditional approaches rely on experimental configurations that yield uniform states of stress and strain (rate), and can therefore be easily inverted to characterize the constitutive law.  However, these approaches only probe highly selected and idealized states and trajectories, thereby leading to large uncertainties in the complex states and trajectories that arise in the applications of the resulting models.  Further, some of these configurations require careful alignment to achieve the idealized states, further limiting the amount of data that can be collected.  In this work, we have introduced an alternate approach that uses a combination of PDE-constrained optimization and full-field observation techniques to obtain material models from complex experiments.  

We formulate the problem of learning the constitutive model as an indirect inverse problem.   We assume that the model is given in a parametrized form: this can be either classical where one has a few parameters or a hyperparametrized formulation like a neural network.  We recast this inverse problem as a PDE-constrained optimization where we seek to find the parameters that minimize the difference between experimentally measured quantities, and the corresponding quantities computed using the parameters subject to the balance laws (PDE-constraint).  We follow an iterative gradient-based approach where we use the adjoint method to compute the sensitivities.  This results in an adjoint partial differential equation (linear in space and quasilinear in time) that is no more expensive to solve than the forward problem.  The resulting algorithm is shown in Figure~\ref{fig:schematic_full_flow}.   As noted in the introduction, our work builds on prior work in the literature on the finite-element updating and PDE-constrained optimization.

In this paper, we describe the formulation, first for a general history dependent material and then for the specific model of a J2 elasto-viscoplasticity.  We then demonstrate it with synthetic data in two experimental configurations: quasistatic compression of a thick plate with a hole, and extended dynamic compression.   In forthcoming work, we demonstrate the method against experimental data (part 2).

We point out a number of salient features of this approach.  First, the method scales linearly (see Section~\ref{sec:scaling}) with the number of parameters in a model with a small pre-factor to the linear term.  Therefore, this method is ideally suited for complex models with a large number of parameters including a hyperparametrized formulation like a neural network.  We demonstrate this in forthcoming work (part 3).  

Second, the method is able to recover complex behavior with a relatively little data.   In the quasistatic test described in Section~\ref{sec:syn_qs}, we only used the strain field at the final time instead of a series of snapshots.  This is because this test used a complex geometry and full-field information, and thus probed multiple trajectories through the adjoint method.  At the same time, the computational cost of the method is relatively independent of the amount of data: so we could have easily incorporated multiple snapshots.

Third, the approach is versatile.  We have demonstrated this by applying it to two different configurations.  Further, the approach is capable of addressing non-smooth problems.  We applied the method here to elasto-viscoplasticity including a yield criterion.  This is difficult to do with a method that uses a numerical derivative.  In forthcoming work (part 2), we extend the approach to indentation including contact.

Finally, the formulation does not make any assumptions about isotropy or homogeneity.  Our current work addresses learning single crystal behavior from observations on polycrystals.

\section*{Acknowledgement}
We are delighted to acknowledge many useful discussions with Professors Ravi Ravichandran and Andrew Stuart.  We are grateful for the financial support of the Army Research Laboratory (W911NF22-2-0120) and the Army Research Office (W911NF-22-1-0269).

\bibliography{Indirect_Inversion}

\appendix
\section{Adjoint problem in elasto-viscoplascitity} \label{app:adj}

We consider an objective of integral from 
    \begin{equation}
    \mathcal{O}(u, q, \varepsilon^p, P) = \int_0^T   \int_{\Omega} o(u, q, \varepsilon^p, P) \, d\Omega .
    \end{equation}
We follow Section \ref{sec:adj} to use the adjoint method to compute the sensitivity of $\mathcal{O}$ with respect to the parameters $P$.  So we augment to objective using the governing equations
    \begin{equation}
    \begin{aligned}
    \mathcal{O} = &\int_0^T  \int_{\Omega} \bigg\{ o + \rho \ddot{u} \cdot v + \pdv{W^e}{\varepsilon} \cdot  \nabla v - b \cdot v + \gamma \dot{q} \left[ \sigma_M - \pdv{W^p}{q} - \pdv{\bar{g}^*}{\dot{q}} \right] + \zeta \cdot \left( \dot{\varepsilon}^p - \dot{q} M \right) \bigg\} \, d\Omega  dt \\
    & + \int_0^T \int_{\partial \Omega} \left( f \cdot{v} \right) \, dS  \, dt.
    \end{aligned}
    \end{equation}
 where the fields $v$, $\gamma$ and $\zeta$  which correspond to the displacement, plastic hardening and plastic strain, respectively are to be determined.   Also, we have used the Kuhn-Tucker condition to replace (\ref{eq:dyn_evo})$_2$ to include the irreversibility of the accumulated plastic strain.  We differentiate this augmented objective with respect to the parameters $P$,
 \begin{equation} \label{eq:long_adj}
    \begin{aligned}
    	\dv{\mathcal{O}}{P} &= \int_{0}^T  \int_{\Omega} \bigg\{ \pdv{o}{P} + \pdv{\rho}{P} \ddot{u} \cdot v + \pdv{^2 W^e}{\varepsilon \partial P} \cdot \nabla v  + \gamma \dot{q} \left( \pdv{\bar{\sigma}_M}{P} - \pdv{^2 W^p}{q \partial P} - \pdv{^2 {g}^*}{\dot{q} \partial P}\right)  \\
    	& \quad + \pdv{o}{u} \delta_P u + \rho v \cdot \delta_P \ddot{u} + \left( \nabla v \cdot \pdv{^2 W^e}{\varepsilon \partial \varepsilon}  +  \gamma \dot{q} \pdv{\bar{\sigma}_M}{\varepsilon} - \dot{q} \zeta \cdot \pdv{M}{\varepsilon} \right) \cdot \nabla \delta_P u  \\
    	& \quad + \left( \pdv{o}{q}  - \gamma \dot{q} \pdv{^2W^p}{q^2}   \right) \delta_P q +   \left( - \gamma \dot{q} \pdv{^2\bar{g}^*}{\dot{q}^2} + \gamma  \left[ \sigma_M - \pdv{W^p}{q} - \pdv{\bar{g}^*}{\dot{q}} \right] - \zeta \cdot M  \right) \delta_P \dot{q} \\
    	& \quad + \zeta \cdot \delta_P \dot{\varepsilon}^p \\
    	& \quad + \left( \pdv{o}{\varepsilon^p} + \nabla v \cdot  \pdv{^2 W^e}{\varepsilon \partial \varepsilon^p }  + \gamma \dot{q}  \pdv{\bar{\sigma}_M}{\varepsilon^p} - \dot{q} \zeta \cdot \pdv{M}{\varepsilon^p} \right) \cdot  \delta_P \varepsilon^p  \bigg \} \ d\Omega \ dt,
    \end{aligned}
    \end{equation}
We then integrate by parts, and enforce quiescent conditions on $v$ at the final time $T$ to remove the boundary terms. Then, localizing gives the sensitivities as
    \begin{equation} \label{eq:sensitivities_app}
    	\begin{aligned}
    		\dv{\mathcal{O}}{P}  = &\int_{0}^T  \int_{\Omega} \bigg\{ \pdv{o}{P} + \pdv{\rho}{P} \ddot{u} \cdot v + \pdv{^2 W^e}{\varepsilon \partial P} \cdot \nabla v + \gamma \dot{q} \left( \pdv{\bar{\sigma}_M}{P} - \pdv{^2 W^p}{q\partial P} - \pdv{^2 {g}^*}{\dot{q} \partial P}\right) \bigg\}  \, d\Omega \, dt,
    	\end{aligned}
    \end{equation}
    if the adjoint variables satisfy the evolution
	\begin{equation} \label{eq:adjoint_system_continuous_app}
    \begin{aligned}
    &0 = \int_{\Omega} \left[ \rho \ddot{v} \cdot \delta_P u + \pdv{o}{u} \cdot \delta_P u  + \left( \nabla v \cdot \pdv{^2 W^e}{\varepsilon \partial \varepsilon}  +  \gamma \dot{q} \pdv{\bar{\sigma}_M}{\varepsilon} - \dot{q} \zeta \cdot \pdv{M}{\varepsilon} \right) \cdot \nabla \delta_P u \right] \ d\Omega && \forall \delta_P u \in \mathcal{K}_0, \\
    & \dv{}{t} \left[ \gamma \left( \bar{\sigma}_M - \pdv{W^p}{q} - \pdv{\bar{g}^*}{\dot{q}}\right)- \gamma \dot{q} \pdv{^2 \bar{g}^*}{\dot{q}^2} - \zeta \cdot M \right]  =\pdv{o}{q} - \gamma \dot{q}  \pdv{^2 W^p}{q^2} && \text{ on } \Omega, \\
    & \dv{\zeta}{t} = \pdv{o}{\varepsilon^p} + \nabla v \cdot \pdv{^2 W^e}{\varepsilon \partial \varepsilon^p} + \gamma \dot{q} \pdv{\bar{\sigma}_M}{\varepsilon^p} - \dot{q} \zeta \cdot \pdv{M}{\varepsilon^p} && \text{ on } \Omega, \\
    & \quad v |_{t = T} = 0, \quad  \dot{v} |_{t = T} = 0, \quad \gamma |_{t = T} = 0, \quad  \zeta |_{t = T} = 0.
    \end{aligned}
    \end{equation}
where $\mathcal{K}_0 := \{ \varphi \in H^1(\Omega), \ \varphi = 0 \text{ on } \partial_u \Omega \}$ is the space of kinematically admissible displacement variations.  This is the weak form of (\ref{eq:vpadj}).

\section{Numerical method in elasto-viscoplasticity} \label{ap:a1}

We discuss the finite element discretization and numerical method to solve the forward and adjoint problems for both the quasistatic and dynamic setting.   We consider a spatial discretization with standard $p = 1$ Lagrange polynomial shape functions for the displacements
\begin{equation}\label{eq:dis_u}
    u = \sum_{i = 1}^{n_u} u_i N_i(x),
\end{equation}
where $N_i : \Omega \mapsto \mathbb{R}^n$ are the standard vector valued shape functions with compact support. The fields $q$ and $\varepsilon^p$ are discretized at quadrature points as
\begin{equation} \label{eq:dis_plas}
    q(x_g) = q_g, \qquad \varepsilon^p(x_g) = \varepsilon^p_g,
\end{equation}
for some Gauss point $x_g$, $g = 1,\dots,n_g$.

\subsection{Forward Problem}

\subsubsection{Quasistatic} \label{ap:forward_qs}
The governing relations in this setting are 
\begin{equation} \label{eq:qs_evo}
    \begin{aligned}
    0 &= \int_\Omega \left[ \mathbb{C} \varepsilon^e \cdot \nabla \delta u - b \cdot \delta u \right] \ d\Omega - \int_{\partial_f \Omega} f \cdot \delta u \ dS \qquad  && \forall \delta u \in \mathcal{K}_0, \\
    0 &\in \sigma_M - \pdv{W^p}{q} - \partial \psi (\dot{q}) \quad && \text{ on } \Omega, \\
    \dot{\varepsilon}^p &= \dot{q} M  && \text{ on } \Omega, \\
    & \hspace{-1em}q|_{t = 0} = 0,\ \varepsilon^p|_{t = 0} = 0 &&
    \end{aligned}
\end{equation}
where $\mathcal{K}_0 := \{ \varphi \in H^1(\Omega), \ \varphi = 0 \text{ on } \partial_u \Omega \}$ is the space of kinematically admissible displacement variations.

We consider a fully implicit approach, with a backwards Euler approximation of the temporal derivatives of the plastic variables. We examine this from the $n$ to $n + 1$ time step. That is, we look to solve for $\{ u^{n+1}, q^{n+1}, \varepsilon^{p,n+1} \}$ given $\{ u^{n}, q^{n}, \varepsilon^{p,n} \}$ assuming the discretizations in~\eqref{eq:dis_u} and~\eqref{eq:dis_plas}. Thus, we look to solve
\begin{equation} \label{eq:dis_qs_evo}
    \begin{aligned}
    0 = R_i &:= \int_\Omega \left[ \mathbb{C} \left( \varepsilon (u^{n+1}) - \varepsilon^{p,n+1} \right) \cdot \nabla N_i - b \cdot N_i \right] \ d\Omega - \int_{\partial_f \Omega} f \cdot N_i \ dS \quad  && i = 1,\dots,n_u \\
    0 &\in \left[ \sigma_M(\nabla u^{n+1}, \varepsilon^{p,n+1}) - \pdv{W^p}{q}(q^{n+1}) - \partial \psi \left( \frac{q^{n+1} - q^n}{\Delta t} \right) \right ]_{x_g} \quad  &&g = 1,\dots,n_g \\
    \left[ \frac{\varepsilon^{p,n+1} - \varepsilon^{p,n}}{\Delta t} \right]_{x_g} &= \left[ \frac{q^{n+1} - q^n}{\Delta t} M(\nabla u^{n+1}, \varepsilon^{p,n+1}) \right]_{x_g} && g = 1,\dots,n_g.
    \end{aligned}
\end{equation}
We solve this through a nested Newton-Raphson approach. From the unidirectional nature of $M$, we have $M(\nabla u^{n+1}, \varepsilon^{p,n+1}) = M(\nabla u^{n+1}, \varepsilon^{p,n})$. Then, from the last line~(\ref{eq:dis_qs_evo}), we may explicitly write $\varepsilon^{p, n+1} = \varepsilon^{p,n+1}(\nabla u^{n+1}, q^{n+1}, \varepsilon^{p, n}, q^n)$. We may then reduce the plastic updates to a single scalar yield equation from the second line of~(\ref{eq:dis_qs_evo}),
\begin{equation} \label{eq:ap_yield_red}
    0 \in \left[ \sigma_M(\nabla u^{n+1}, \varepsilon^{p,n+1}(\nabla u^{n+1}, q^{n+1}, \varepsilon^{p, n}, q^n)) - \pdv{W^p}{q} \  (q^{n+1}) - \partial \psi \left( \frac{q^{n+1} - q^n}{\Delta t} \right) \right ]_{x_g}
\end{equation}
for $g = 1,\dots,n_g$. Then, given $\nabla u^{n+1}$, $\varepsilon^{p,n}$, and $q^n$, the above equation is a scalar relation for $q^{n+1}$ at each quadrature point. We solve this though a standard Newton-Raphson method. Then, from the solution of the above, we may write $q^{n+1} = q^{n+1}(\nabla u^{n+1}, \varepsilon^{p, n}, q^n)$ and therefore $\varepsilon^{p,n+1} = \varepsilon^{p,n+1}(\nabla u^{n+1}, \varepsilon^{p,n}, q^n)$. This allows us to reduce the entire system to
\begin{equation}
    0 = R_i := \int_\Omega \left[ \mathbb{C} \left( \varepsilon (u^{n+1}) - \varepsilon^{p,n+1}(\nabla u^{n+1}, \varepsilon^{p, n}, q^n)  \right) \cdot \nabla N_i - b \cdot N_i \right] \ d\Omega - \int_{\partial_f \Omega} f \cdot N_i \ dS 
\end{equation}
for $i = 1,\dots,n_u$, where the plastic update relations are accounted for through the dependence of $\varepsilon^{p,n+1}$ on $\nabla u^{n+1}, \ \varepsilon^{p,n}$ and $q^n$. We solve this system through a Newton-Raphson method. Thus, it is necessary to compute the stiffness matrix
\begin{equation}
    K_{ij} := \int_\Omega \left[  \mathbb{C} \left( \nabla N_j - \pdv{\varepsilon^{p,n+1}}{\nabla u} \cdot \nabla N_j  \right) \cdot \nabla N_i \right] ] \ d\Omega 
\end{equation}
where 
\begin{equation}
    \pdv{\varepsilon^{p,n+1}}{\nabla u} = M(\nabla u^{n+1}, \varepsilon^{p,n}) \otimes \pdv{q^{n+1}}{\nabla u} + (q^{n+1} - q^n) \pdv{M}{\nabla u}. 
\end{equation}
Here, the derivative $\pdv{q^{n+1}}{\nabla u}$ is found through an implicit differentiation of~\eqref{eq:ap_yield_red} giving
\begin{equation}
    \pdv{q^{n+1}}{\nabla u} = \frac{2 \mu}{\pdv{^2 W^p}{q^2} + \frac{1}{\Delta t} \pdv{^2 \psi}{\dot{q}^2} + 3 \mu} M(\nabla u^{n+1}, \varepsilon^{p,n}) .
\end{equation}

Then, starting from $u^{n+1, 1} = u^n$ and $p = 1$, we conduct Newton Raphson iterations over $p$ in the form
\begin{equation}
\begin{aligned}
    K_{ij}(u^{n+1, p}) \Delta u_j &= - R_i(u^{n+1, p}), \quad && i = 1,\dots,n_u, \ j=1,\dots,n_u\\
    u_j^{n+1, p+1} &= \Delta u_j + u_j^{n+1, p}, &&  j = 1,\dots,n_u.  
\end{aligned}
\end{equation}
until $|R(u^{n+1, p+1})| < tol$, and set $u^{n+1} = u^{n+1, p+1}$ and $q^{n+1} = q^{n+1}(\nabla u^{n+1}, \varepsilon^{p, n}, q^n)$ and $\varepsilon^{p,n+1} = \varepsilon^{p,n+1}(\nabla u^{n+1}, \varepsilon^{p,n}, q^n)$.

\subsubsection{Dynamic}


We specialize to the dynamic compression test where a disc (possibly with a hole) is compressed axially.  See Section {\ref{sec:syn_dyn}) and Appendix (\ref{sec:annulus}).  The governing equations are (\ref{eq:append_B_dyn_evo_02}).

We use an explicit central difference scheme to update the displacement field. The plasticity updates $q$ and $\varepsilon^p$ are then updated implicitly with a backwards Euler update. For the $n$ to $n+1$ time-step the displacement updates are
\begin{equation}
\begin{aligned}
\ddot{\bar{u}}_i^{n} &= M^{-1}_{ij} F_j^n(\bar{u}^n, \varepsilon^{p,n}, q^n, t^n),\\
\dot{\bar{u}}_i^{n + 1/2} &= \dot{\bar{u}}_i^{n - 1/2} + \Delta t^n \, \ddot{\bar{u}}_i^{n}, \\
\bar{u}_i^{n + 1} &= \bar{u}_i^n + \Delta t^{n + 1/2} \, \dot{\bar{u}}_i^{n + 1/2},
\end{aligned}
\end{equation}
where
\begin{equation}
M_{ij} = \int_{\Omega} \rho(x) N_i \cdot N_i \, d\Omega, \qquad F_j^n = \int_{\Omega} \left[ -\sigma (\varepsilon^n, \varepsilon^{p,n},\alpha^n, \eta) \cdot \nabla N_j  + b \cdot N_j \right] \, d\Omega.
\end{equation}
In standard fashion, the integrals above are approximated with Gauss quadrature. We then update the plasticity variables through an implicit backwards Euler discretization. For this, we employ a predictor-corrector scheme~\cite{Ortiz1999} to solve point-wise at each quadrature point,
\begin{equation}
\begin{aligned}
& 0 \in \bar{\sigma}_M (\varepsilon^{n + 1} |_{x_g}, \varepsilon^{p, (n + 1)}_g, \eta(x_g) ) - \pdv{W^p}{q}(q^{n+1}_g, \eta(x_g) ) - \partial g^* \left( \frac{q^{n + 1}_g - q^n_g}{\Delta t}, \eta(x_g) \right), \\
& \varepsilon^{p, (n+1)}_g = \varepsilon^{p, n}_g + \Delta q M (\varepsilon^{n + 1}_g, \varepsilon^{p, (n + 1)}_g).
\end{aligned}
\end{equation} 

\subsection{Adjoint Problem } 
%
\subsubsection{Quasistatic}  \label{ap:adjoint_qs}

The governing set of quasistatic adjoint relations are
\begin{equation} \label{eq:adjoint_system_qs_app}
    \begin{aligned}
    &0 = \int_{\Omega} \left[\pdv{o}{u} \cdot \delta_P u  + \left( \nabla v \cdot \pdv{^2 W^e}{\varepsilon \partial \varepsilon}  +  \gamma \dot{q} \pdv{\bar{\sigma}_M}{\varepsilon} - \dot{q} \zeta \cdot \pdv{M}{\varepsilon} \right) \cdot \nabla \delta_P u \right] \ d\Omega && \forall \delta_P u \in \mathcal{U}, \\
    & \dv{}{t} \left[ \gamma \left( \bar{\sigma}_M - \pdv{W^p}{q} - \pdv{\bar{g}^*}{\dot{q}}\right)- \gamma \dot{q} \pdv{^2 \bar{g}^*}{\dot{q}^2} - \zeta \cdot M \right]  =\pdv{o}{q} - \gamma \dot{q}  \pdv{^2W^p}{q^2} && \text{ on } \Omega, \\
    & \dv{\zeta}{t} = \pdv{o}{\varepsilon^p} + \nabla v \cdot \pdv{^2 W^e}{\varepsilon \partial \varepsilon^p} + \gamma \dot{q} \pdv{\bar{\sigma}_M}{\varepsilon^p} - \dot{q} \zeta \cdot \pdv{M}{\varepsilon^p} && \text{ on } \Omega, \\
    & \quad  \gamma |_{t = T} = 0, \quad  \zeta |_{t = T} = 0.
    \end{aligned}
\end{equation}

As the boundary conditions are found at the final time $t = T$, we solve this system backwards in time. We consider a fully implicit approach, with a backward Euler approximation of the temporal derivatives of the plastic variables. We examine this from the $n+1$ to $n$ time step. That is, we look to solve for $\{ v^{n}, \gamma^{n}, \zeta^{n} \}$ given $\{ v^{n+1}, \gamma^{n+1}, \zeta^{n+1} \}$ assuming the discretizations in~\eqref{eq:dis_u} and~\eqref{eq:dis_plas} given the complete set of forward-problem solution variables. Thus, we look to solve
\begin{equation} \label{eq:adjoint_system_qs_app_disc}
    \begin{aligned}
    &0 = R^{adj}_i =  \int_{\Omega} \left[ \left. \pdv{o}{u} \right |_{t_n} \cdot N_i  + \left( \nabla v^n \cdot \left. \pdv{^2 W^e}{\varepsilon \partial \varepsilon} \right |_{t_n}  +  \gamma^n \left( \dot{q} \pdv{\bar{\sigma}_M}{\varepsilon} \right )_{t_n} - \zeta^n \cdot \left ( \dot{q} \pdv{M}{\varepsilon}  \right )_{t_n} \right) \cdot \nabla N_i \right] \ d\Omega  \\
    & \frac{1}{\Delta t } \left[ \gamma \left( \bar{\sigma}_M - \pdv{W^p}{q} - \pdv{\bar{g}^*}{\dot{q}}\right)- \gamma \dot{q} \pdv{^2 \bar{g}^*}{\dot{q}^2} - \zeta \cdot M \right]^{t_{n+1}}_{t_n}  = \left. \pdv{o}{q} \right |_{t_n} - \gamma^n \left( \dot{q}  \pdv{^2W^p}{q^2} \right )_{t_n}\\
    & \frac{\zeta^{n+1} - \zeta_n}{\Delta t} = \left . \pdv{o}{\varepsilon^p} \right |_{t_n} + \nabla v^n \cdot \left . \pdv{^2 W^e}{\varepsilon \partial \varepsilon^p} \right |_{t_n} + \gamma^n \left( \dot{q} \pdv{\bar{\sigma}_M}{\varepsilon^p} \right)_{t_n} -  \zeta^n \cdot \left( \dot{q} \pdv{M}{\varepsilon^p} \right)_{t_n}
    \end{aligned}
\end{equation}
This is a linear set of equations for the  $\{ v^{n}, \gamma^{n}, \zeta^{n} \}$ and can be solved through direct inversion.

\subsubsection{Dynamic}

Similar to the forward problem, we apply an explicit central difference scheme to update the adjoint field $v$. The other adjoint variables, namely $\gamma$  and $\zeta$ are solved locally using  then updated implicitly with a backwards Euler update. For the $n$ to $n+1$ time-step the displacement updates are
\begin{equation}
\begin{aligned}
\ddot{\bar{v}}_i^{n} &= M^{-1}_{ij} H_j^n(\bar{u}^n, \varepsilon^{p,n}, q^n, t^n),\\
\dot{\bar{v}}_i^{n + 1/2} &= \dot{\bar{v}}_i^{n - 1/2} + \Delta t^n \, \ddot{\bar{v}}_i^{n}, \\
\bar{v}_i^{n + 1} &= \bar{v}_i^n + \Delta t^{n + 1/2} \, \dot{\bar{v}}_i^{n + 1/2},
\end{aligned}
\end{equation}
where
\begin{align}
M_{ij} &= \int_{\Omega} \rho(x) N_i \cdot N_i \, d\Omega, \nonumber \\  
H_j^n &= \int_{\Omega} \left[ - \left( \nabla v^n \cdot \pdv{^2 W^e}{\varepsilon \partial \varepsilon}  +  \gamma^n \dot{q}^n \pdv{\bar{\sigma}_M}{\varepsilon} - \dot{q}^n \zeta^n \cdot \pdv{M}{\varepsilon}\right) \cdot \nabla N_j  - \frac{\partial o}{\partial u} \cdot N_j \right] \, d\Omega.
\end{align}
In standard fashion, the integrals above are approximated with Gauss quadrature. We then update the adjoint variables $\gamma$ and $\zeta$ through an implicit backwards Euler discretization. For this, we solve a linear set of equations point-wise at each quadrature point,
\begin{equation}
\begin{aligned}
& \gamma^{n+1} \left( \bar{\sigma}_M - \pdv{W^p}{q} - \pdv{\bar{g}^*}{\dot{q}} -  \dot{q} \pdv{^2 \bar{g}^*}{\dot{q}^2}\right)
- \zeta^{n+1} \cdot M 
= \gamma^{n} \left( \bar{\sigma}_M - \pdv{W^p}{q} - \pdv{\bar{g}^*}{\dot{q}} -  \dot{q} \pdv{^2 \bar{g}^*}{\dot{q}^2} \right), \\
&  - \zeta^{n} \cdot M + \Delta t \frac{\partial o}{\partial q} - \Delta t \gamma^n \dot{q} \pdv{^2W^p}{q^2} , \\ 
& \zeta^{n+1}   = \zeta^n +  \pdv{o}{\varepsilon^p} + \nabla v \cdot \pdv{^2 W^e}{\varepsilon \partial \varepsilon^p} + \gamma ^n\dot{q} \pdv{\bar{\sigma}_M}{\varepsilon^p} - \dot{q} \zeta^n \cdot \pdv{M}{\varepsilon^p}.
\end{aligned}
\end{equation}

\section{Dimension Reduction for Annular Specimen}
\label{sec:annulus}

The annulus ring being thin allows us to perform computations on a reduced dimensional space enabling quicker algorithm. Reiterating our approximation of uniform axial strain along the $e_3$ direction, 
\begin{equation}  \label{eq:append_B_u_assump}
    u (X_1, X_2, X_3, t) = \bar{u}(X_1, X_2, t) + (\lambda(t)-1) X_3 \ e_3,
\end{equation}
where $\bar{u} := \bar{\Omega} \mapsto \mathbb{R}^2$ is the in-plane displacement. The corresponding deformation gradient is given by
\begin{align}
    \nabla u = \begin{pmatrix}
    \bar{\nabla}\bar{u} & 0 \\ 
    0 & \lambda
    \end{pmatrix},
\end{align}
where $\bar{\nabla}$ is the gradient computed along the in-plane directions $\{ e_1, e_2\}$. The governing equations are,
\begin{equation} \label{eq:dyn_evo_red}
    \begin{aligned}
    0 &= \int_\Omega \left[\rho \ddot{u} \cdot \delta u +   \sigma \cdot \nabla \delta u - b \cdot \delta u \right] \ d\Omega - \int_{\partial_f \Omega} f \cdot \delta u \ dS \qquad  && \forall \delta u \in \mathcal{K}_0, \\
    0 &\in \sigma_M - \pdv{W^p}{q} - \partial \psi (\dot{q}) \quad && \text{ on } \Omega, \\
    \dot{\varepsilon}^p &= \dot{q} M  && \text{ on } \Omega.
    \end{aligned}
\end{equation}
Substituting the approximation of the displacement \ref{eq:append_B_u_assump} into the above equations, we obtain the simplified form 
\begin{equation} \label{eq:append_B_dyn_evo_02}
    \begin{aligned}
    0 &= -\frac{h^2 ||\Omega||}{3}\rho \ddot{\lambda} - \int_\Omega \sigma_{33} + \int_{\Omega} \ f \cdot e_3 \ ds \ \Big \vert_{X_3 = h}, \\
    0 &= \int_\Omega \left[\rho \ddot{\bar{u}} \cdot \delta \bar{u} +   \sigma \cdot \nabla \delta \bar{u} - b \cdot \delta \bar{u} \right] \ d\Omega  \qquad  && \forall \delta \bar{u} \in \mathcal{K}_0, \\
    0 &\in \sigma_M - \pdv{W^p}{q} - \partial \psi (\dot{q}) \quad && \text{ on } \Omega, \\
    \dot{\varepsilon}^p &= \dot{q} M  && \text{ on } \Omega.
    \end{aligned}
\end{equation}
Since $\bar{u}$ exists in a 2-dimensional space, the computational cost for solving the above equations are cheaper than solving for a 3-dimensional system of equations. The dynamic boundary condition involves specifying $\lambda (t)$; therefor, the first equation is not necessary. The net force on the surface $X_3=h$ can be calculated using 
\begin{align} \label{eq:append_B_surface_force}
    f = \int_\Omega \sigma_{33} \ d \Omega \ \Big \vert_{X_3 = h}
\end{align}
Our experimental data consists of the final in plane displacement $\bar{u}^\text{exp}(X_1,X_2,T)$ and the net axial force $f_R^\text{exp}$, which are compared with the inplane-displacements obtained from equation \ref{eq:append_B_dyn_evo_02} and the surface forces computed using \ref{eq:append_B_surface_force}.

%
%

\end{document}